\documentclass[pra,onecolumn,floatfix,a4paper,superscriptaddress]{revtex4}
\usepackage{bm,color,graphicx,amsmath,txfonts}
\usepackage{float}
\usepackage[colorlinks, citecolor=blue,linkcolor=blue]{hyperref}

\newcommand{\diag}{{diag}}

\begin{document}

\title{Magnon-rotation enhanced nonreciprocity of multipartite entanglement in a magnomechanical system}

\author{Hamza Harraf}
\affiliation{LPHE-Modeling and Simulation, Faculty of Sciences, Mohammed V University in Rabat, Rabat, Morocco}
\author{Noura Chabar} 
\affiliation{LPTHE-Department of Physics, Faculty of Sciences, Ibnou Zohr University, Agadir 80000, Morocco}	
\author{Mohamed Amazioug} \thanks{m.amazioug@uiz.ac.ma}
\affiliation{LPTHE-Department of Physics, Faculty of Sciences, Ibnou Zohr University, Agadir 80000, Morocco}
\author{Rachid Ahl Laamara}
\affiliation{LPHE-Modeling and Simulation, Faculty of Sciences, Mohammed V University in Rabat, Rabat, Morocco}
\affiliation{Centre of Physics and Mathematics, CPM, Faculty of Sciences, Mohammed V University in Rabat, Rabat, Morocco}
\author{Mojtaba Mazaheri}
\affiliation{Department of Basic Science, Hamedan University of Technology, Hamedan 65169-1-3733, Iran}

\begin{abstract} 

Nonreciprocal physics is attracting significant interest in quantum information processing. In this work, we propose a scheme to investigate the nonreciprocity of bi- and tripartite entanglement and generate squeezed states in a magnomechanical system. This is achieved through the Barnett effect, which originates from the rotation of the first magnon mode. The system consists of two YIG spheres, each supporting a magnon mode that represents collective spin motion, positioned inside a microwave cavity (MC). We show that the Barnett effect enhances entanglement under thermal effects and generates squeezed states for the two magnon modes and the photon mode. Moreover, we show that magnon-magnon coupling enhances entanglement between different two modes.

\textbf{Keywords :} Cavity magnomechanics; Barnett effect; Nonreciprocity; Entanglement; Yttrium Iron Garnet (YIG).

\end{abstract}

\date{\today}
\maketitle

\section{Introduction}

Over the last few years, ferromagnetic materials like the yttrium iron garnet (YIG) sphere have garnered significant attention and achieved remarkable progress in the field of quantum information processing \cite{intro1}. Magnons, which are collective spin excitations in the YIG sphere, can couple with microwave cavities strongly and even ultra-strongly due to their high spin density and low damping rates \cite{intro2}. In Cavity Magnomechanics (CM), the magnetostrictive effect, caused by the deformation of the YIG's geometric structure, drives the magnomechanical interaction that couples a YIG sphere's magnon mode to its vibrational mode (phonon) \cite{intro3,ramshti2022}. These characteristics have opened new avenues for exploring intriguing phenomena, such as magnon-induced nonreciprocity \cite{intro4}, quadrature squeezing \cite{intro5}, and magnon blockade \cite{intro6}. Recent research in cavity magnomechanics has placed particular emphasis on the preparation of macroscopic entangled states, given their crucial role not only in quantum information processing but also in probing fundamental aspects of quantum theory. For example, Li et al. put forth a simple and workable plan for producing genuine tripartite entanglement between magnons, photons, and phonons \cite{intro7}, as well as bipartite entanglement. Subsequent studies have also demonstrated the realization of entanglement in hybrid multi-mode CM systems, including entanglement between distinct magnon modes \cite{intro8}, photon-magnon entanglement \cite{intro9}, and atom-magnon entanglement \cite{intro10}.\\

Quantum entanglement, one of the fundamental concepts of quantum physics, describes the inseparability of particles. This nonclassical correlation is a primary distinction between quantum physics and classical mechanics. Because of this unique characteristic, entanglement is frequently used in quantum communication \cite{intro11}, quantum teleportation \cite{intro12}, quantum computation \cite{intro13}, and many other fields. In recent years, many techniques have been proposed to create steady-state entanglement in a variety of quantum systems, including reservoir engineering \cite{intro14}, dark mode engineering \cite{intro15}, and coherent feedback \cite{intro16}. Furthermore, microwave-optical entanglement based on magnons \cite{intro17} and tripartite entanglement between a phonon, a magnon, and a photon \cite{intro18} have also been implemented in cavity magnomechanical systems \cite{zuo2024}. However, unpredictable interactions between quantum systems and the environment make it difficult to prepare long-lived entangled states, particularly on the macroscopic scale. As a result, there is great interest in attempts to create macroscopic entangled states using different physical setups. In this context, steady-state entanglement between atomic ensembles has been successfully demonstrated \cite{intro19}. Furthermore, entangled states involving macroscopic systems have been documented in many configurations. For example, entanglement has been shown between a mechanical resonator and a single cavity mode in an optomechanical arrangement \cite{intro20}. Recently, entanglement between two macroscopic resonators connected to a common cavity mode via optomechanical interaction has been experimentally achieved \cite{intro21}.\\

Nonreciprocal entanglement was initially proposed in a spinning resonator \cite{70c,71c}, which generates an irreversible refractive index for the clockwise and counterclockwise modes. This leads to the breaking of time-reversal symmetry within the system, enabling the entanglement of photons and phonons in a designated direction while significantly suppressing entanglement in the opposite direction. Additionally, nonreciprocal entanglement among different modes has been investigated based on the Sagnac effect \cite{72c,73c,74c}, magnon Kerr nonlinearity \cite{75c,76c}, chiral coupling \cite{77c,78c}, and photon and magnon blockade \cite{amazioug2025}.\\

In this work, we investigate the nonreciprocity of bi- and tripartite entanglement and generate squeezed states of magnons and photons in a magnomechanical system. This is accomplished by implementing the Barnett effect, which improves upon the results obtained in \cite{intro8}. The Barnett effect has been noted in ferromagnetic insulators, where the rotation of an object with magnetic moments results in alignment or magnetization \cite{kani2022}, and also in nuclear spin systems \cite{intro23}. This phenomenon has been used to remotely investigate switch magnetization \cite{intro24}, explore rotational vacuum friction \cite{intro25}, and detect the angular momentum compensation point \cite{intro26}. The Barnett effect (BE) refers to the magnetization induced in a magnetic material such as a YIG sample, when it undergoes rapid rotation. This rotation leads to a shift in the Barnett frequency, which can be tuned to positive or negative values by reversing the direction of the applied magnetic field \cite{barnett,barnett2,zhang2025,barnett1915}. Additionally, we show that the Barnett effect enhances entanglement under thermal effects and generates squeezed states for the two magnon modes and the photon mode. Moreover, magnon-magnon coupling enhances entanglement between different two modes.\\

The outline of the paper is as follows. In Section \ref{sec2}, we introduce the theoretical model by presenting the system's Hamiltonian, providing the equations of motion for the system operators, calculating the expectation values of these operators, and obtaining the linearized quantum Langevin equations. Section \ref{sec3} focuses on the linearization of the quantum Langevin equations in the steady state and the calculation of the covariance matrix (CM). In Section \ref{sec4}, we employ logarithmic negativity and minimal residual contangle to quantify bipartite and tripartite entanglement, respectively. In Section \ref{sec6}, we discuss the bipartite, tripartite, and nonreciprocal entanglement under different system parameters. In section \ref{sec7}, we present the concluding remarks that close this paper.

\section{Model}  \label{sec2}

In this model, we propose a hybrid cavity magnomechanical system that consists of a microwave cavity mode ($\mathsf{c}$), a mechanical mode ($\mathsf{b}$), and two magnon modes ($\mathsf{m}_1$ and $\mathsf{m}_2$). The two magnon modes, representing the collective motions of many spins in two macroscopic YIG spheres, couple to a single microwave cavity simultaneously. This connection between photons and magnons is mediated by magnetic dipole interactions. This system, without a mechanical mode and with two YIG spheres, has been previously used to study magnon dark modes \cite{zhang2015magnon}. The mechanical mode here is the vibrational deformation of the YIG crystal, caused by a magnetostrictive force. The type of magnetostrictive interaction is influenced by the resonance frequencies of the phonon and magnon modes \cite{kittel1958interaction}. Generally, the magnon-phonon coupling is weak because the mechanical mode's frequency is much lower than the magnon mode's \cite{zhang2016cavity}. However, applying a strong microwave field significantly enhances this coupling \cite{wang2018bistability}. The direction of the bias magnetic field determines the magnomechanical coupling strength \cite{kittel1958interaction}. We adjust the orientations of the two bias magnetic fields so that the magnetostrictive interaction is effectively activated for only one of the spheres. Next, the system's Hamiltonian is written as ($\hbar=1$)
\begin{equation}
	\begin{aligned}
		\label{eq1}
		\mathcal{H} =& \, \omega_{\mathsf{c}} \mathsf{c}^{\dagger}\mathsf{c}+ (\omega_{\mathsf{m}_1}+\Delta_B)\mathsf{m}^{\dagger}_1 \mathsf{m}_1+\omega_{\mathsf{m}_2}\mathsf{m}^{\dagger}_2 \mathsf{m}_2+\frac{\omega_{\mathsf{b}}}{2}(\mathsf{\hat{y}}^2+\mathsf{\hat{x}}^2)\\
		&+ \sum_{j=1}^2 
		 \mathsf{g_j}(\mathsf{c}+\mathsf{c}^{\dagger})(\mathsf{m}^{\dagger}_j+\mathsf{m}_j)
		+ \mathsf{G}_0 \mathsf{m}_1^{\dagger} \mathsf{m}_1 \mathsf{\hat{x}} 
		+ \mathsf{J}(\mathsf{m}_1^{\dagger}\mathsf{m}_2+\mathsf{m}_1\mathsf{m}_2^{\dagger})+i \Psi \left(\mathsf{m}_1^{\dagger} e^{-i\omega_{0}t} - \mathsf{m}_1 e^{i\omega_{0}t} \right),
	\end{aligned}
\end{equation}
the operators $\mathsf{c}^{\dagger}$ and $\mathsf{c}$ $\left(\mathsf{m}^{\dagger}_{1(2)} \text{ and } \mathsf{m}_{1(2)}\right)$ represent the creation and annihilation operators for the cavity mode and magnon modes, respectively, they satisfy the commutation relation \([S, S^{\dagger}] = 1\), where \(S = \mathtt{c} (\mathsf{m}_{1(2)})\). The dimensionless position and momentum quadratures of the mechanical modes give by \(\mathsf{\hat{x}}\) and \(\mathsf{\hat{y}}\) satisfy \([\mathsf{\hat{x}}, \mathsf{\hat{y}}] = i\). The resonance frequencies of the cavity, magnon, and mechanical modes are denoted by \(\omega_{\mathsf{c}}\), \(\omega_{\mathsf{m}_{1(2)}}\), and \(\omega_{\mathsf{b}}\), respectively. The magnon frequencies, \(\omega_{\mathsf{m}_{1(2)}}\) are established by the external bias magnetic fields \(\mathsf{H}_{1(2)}\) through the relation \(\omega_{\mathsf{m}_{1(2)}} = \Gamma \mathsf{H}_{1(2)}\), where \(\Gamma / 2\pi = 28\) GHz /T represents the gyromagnetic ratio. The angular frequency of the YIG sphere's rotation around the \(z\) axis is \(\Delta_{B}\) considering the Barnett effect \cite{barnett,Barnett1917,ono2015,jia2025}, an emerging magnetic field is produced by a rotating YIG sphere \(\Delta_B=\Gamma\mathsf{H}_B\). Consequently, the magnon(1) frequency $\omega_{\mathsf{m}_1}$ experiences a shift to $\omega_{\mathsf{m}_1}+\Delta_B$ \cite{Ge2025}. The linear coupling between the cavity and the two magnon modes represents by the coupling rate \(\mathsf{g}_{1(2)}\), whereas the single-magnon magnomechanical coupling rate is indicated by \(\mathsf{G}_0\). We denote \(\mathsf{J}\) the coupling strength between two magnon modes.  The intensity of the coupling between the drive magnetic field (with amplitude $\mathsf{B}_0$ and frequency $\omega_0$) and the first magnon mode is measured by the Rabi frequency \(\Psi = \frac{\sqrt{5\mathsf{N}}}{4} \Gamma \mathsf{B}_0\) \cite{li2019entangling}. Here, \(\mathsf{N} = \rho \mathsf{V}\) represents the total number of spins, where \(\rho = 4.22 \times 10^{27} \, 1/\mathrm{m}^{3}\) is the spin density of YIG, and the volume of sphere is \(\mathsf{V}\). The first and second (third) term in the Hamiltonian characterize the energies of the cavity mode and magnon modes, while the fourth term represents the energy of the mechanical vibration mode with frequency \(\omega_{\mathsf{b}}\). The fifth, sixth, and seventh terms represent the cavity magnon coupling, the single-magnon magnomechanical coupling, and the two-magnon coupling, respectively. $\mathsf{J}$ denotes the magnon-magnon coupling strength, which can reach up to the strong-coupling regime by varying the distance between the two YIG spheres \cite{X,Y,Berinyuy2025,zheng2024}. The eighth term describes the driving Hamiltonian. Driving the magnon mode $\mathsf{m}_1$ with a strong microwave field can significantly enhance the effective magnomechanical coupling \cite{li2018magnon}. Note that this model differs from the one presented in Ref. \cite{li2019entangling}. 

\begin{figure*}[!h]
\centering
\includegraphics[width=0.3\linewidth, height=0.3\linewidth]{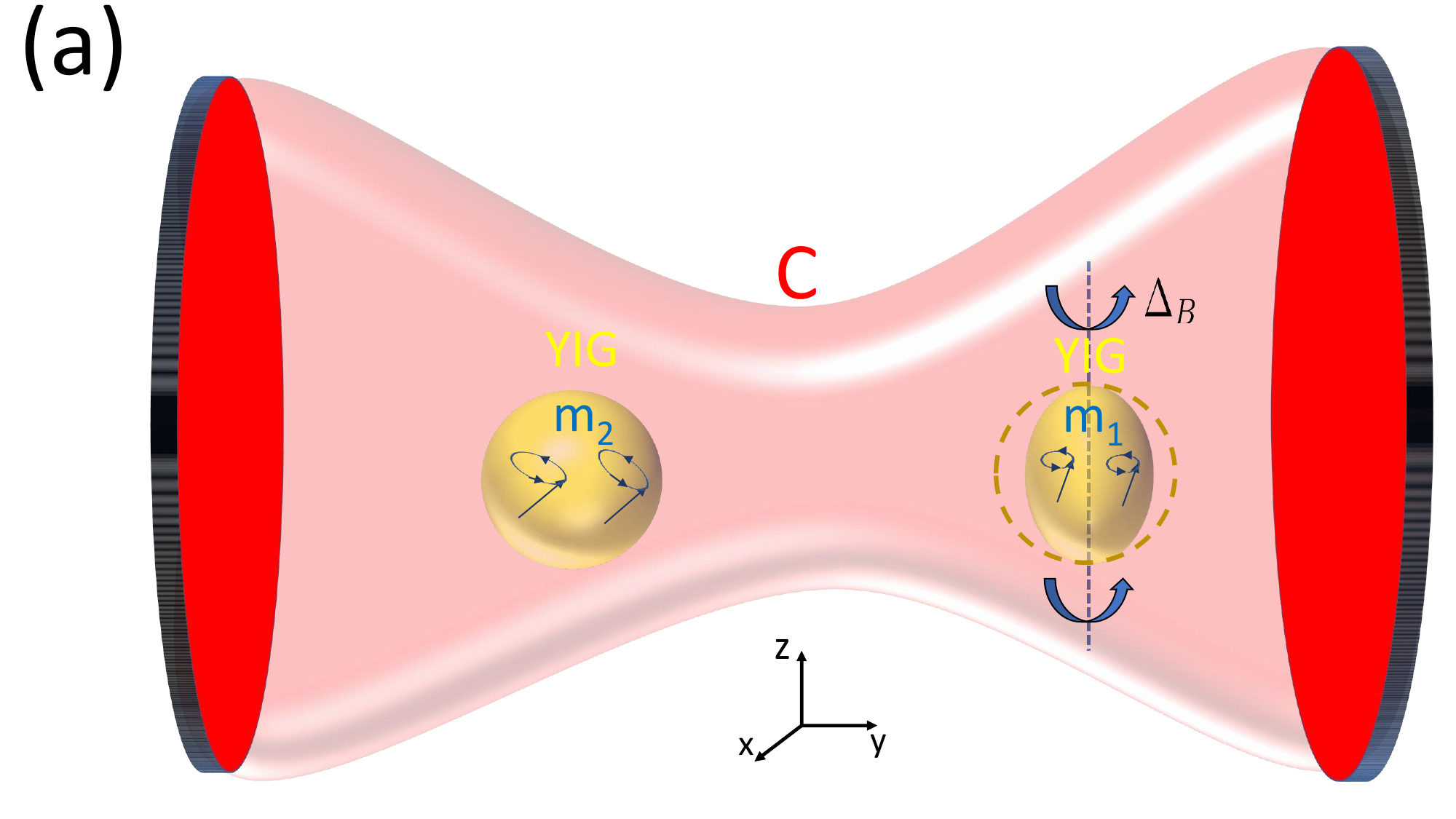}\hspace{0.5cm}
\includegraphics[width=0.3\linewidth, height=0.3\linewidth]{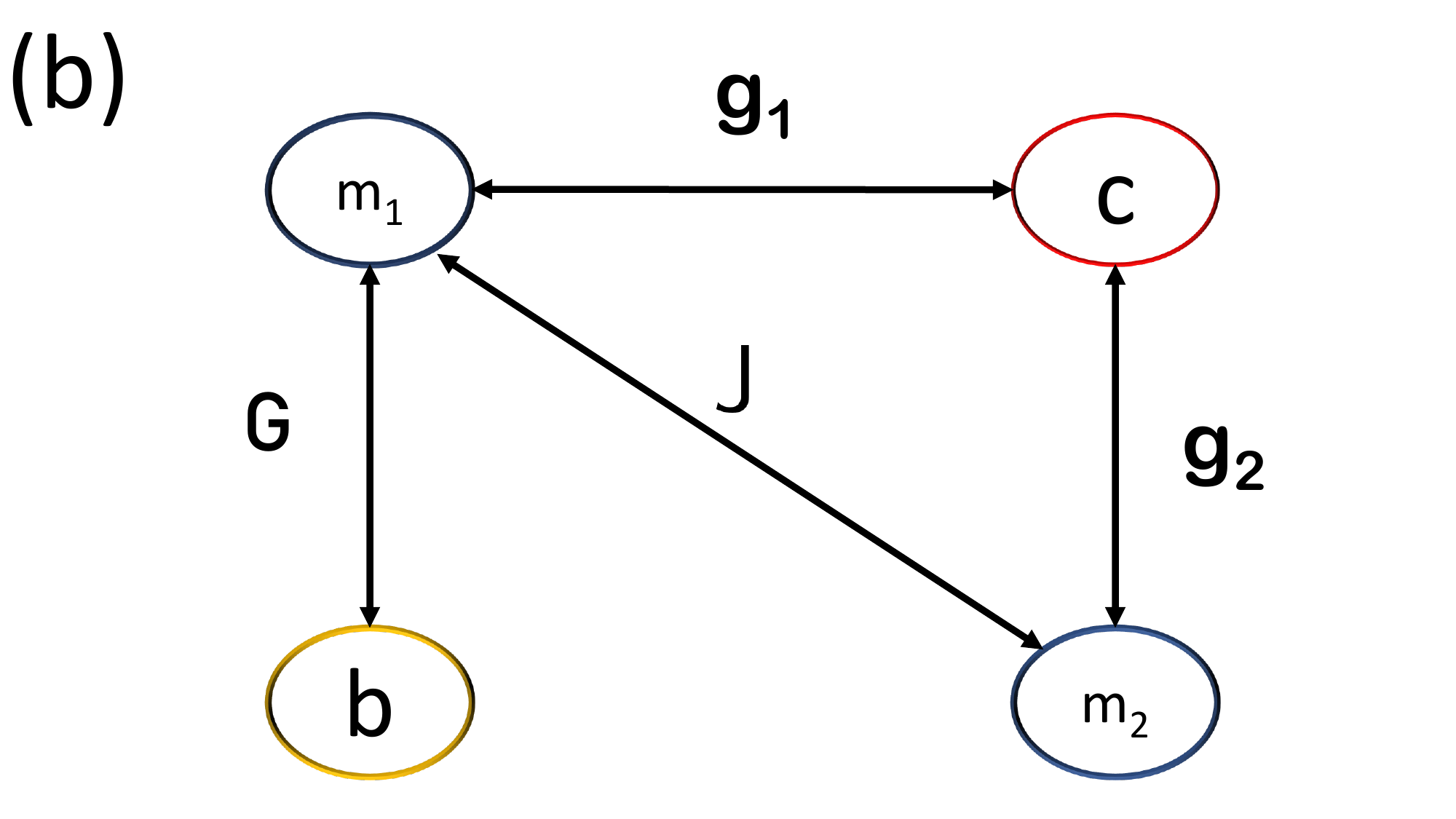}\hspace{0.5cm}
\includegraphics[width=0.3\linewidth, height=0.3\linewidth]{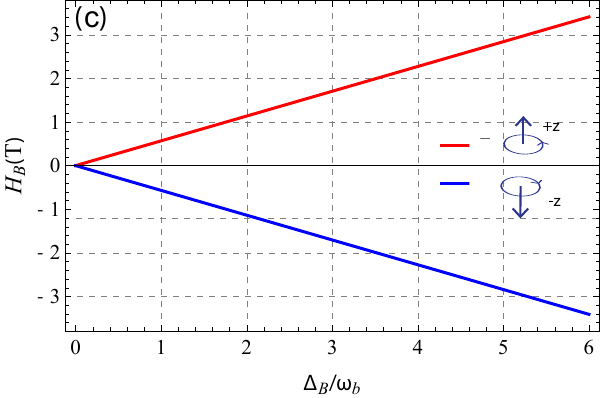}
\caption{(a) This schematic illustrates the hybrid four-mode cavity magnomechanical system. Two YIG samples are positioned within a microwave cavity at points of maximum magnetic field for the cavity mode. They are concurrently exposed to uniform bias magnetic fields that excite magnon modes and couple them to the cavity mode. The bias magnetic fields are oriented to activate the magnetostrictive (magnon-phonon) interaction in only one YIG sample (YIG1). This coupling can be further enhanced by directly driving the magnon mode with an external microwave source (not shown). The rotating YIG sphere with an angular frequency $\Delta_B$ will create an emergent magnetic field $\mathsf{H}_B$ that causes the magnon(1) to experience a frequency shift. (b) The diagram illustrates the constant coupling between the directly interacting modes. The cavity mode is linearly coupled to the two magnon modes, $\mathsf{m}_1$ and $\mathsf{m}_2$, with coupling constants $\mathsf{g}_1$ and $\mathsf{g}_2$, respectively. The two magnon modes are also coupled by a strong coupling $\mathsf{J}$. Additionally, magnon mode $\mathsf{m}_1$ interacts with the mechanical mode $\mathsf{b}$ via a nonlinear magnetostrictive interaction, which is characterized by an effective coupling rate $\mathsf{G}$. This interaction facilitates magnomechanical entanglement \cite{li2018magnon}, which can, in turn, be used to entangle the two magnon modes. Indirect couplings are not depicted in this scheme. (c) A plot of $\mathsf{H}_B$ versus $\Delta_B$ for different rotations of the magnon around the $z$-axis.}
\label{DOC}
\end{figure*}
In the frame rotating at the drive frequency $\omega_0$ and by applying the rotating wave approximation (RWA), the expression $\mathsf{g}_{1(2)}(\mathsf{c}+\mathsf{c}^{\dagger})(\mathsf{m}_{1(2)}+\mathsf{m}_{1(2)})\rightarrow\mathsf{g}_{1(2)}(\mathsf{c}\mathsf{m}_{1(2)}+\mathsf{c}^{\dagger}\mathsf{m}_{1(2)})$, (valid when $\omega_{\mathsf{c}}, \omega_{\mathsf{m}_{1(2)}} \gg \mathsf{g}_{1(2)}, \mathsf{k}_{\mathsf{c}}, \mathsf{k}_{\mathsf{m}}$, which is easily satisfied \cite{zhang2016cavity}, the Hamiltonian is changed to
\begin{equation}
	\begin{aligned}
		\label{eq1}
		\mathcal{H} =& \, \Delta_{\mathsf{c}} \mathsf{c}^{\dagger}\mathsf{c}
		+ (\Delta_{\mathsf{m}_1}+\Delta_{B})\mathsf{m}^{\dagger}_1 \mathsf{m}_1+\Delta_{\mathsf{m}_2}\mathsf{m}^{\dagger}_2 \mathsf{m}_2+\frac{\omega_{\mathsf{b}}}{2}(\mathsf{\hat{y}}^2+\mathsf{\hat{x}}^2)\\
		+ &\sum_{j=1}^2 
		 \mathsf{g_j}(\mathsf{c}+\mathsf{c}^{\dagger})(\mathsf{m}^{\dagger}_j+\mathsf{m}_j)
		+ \mathsf{G}_0 \mathsf{m}_1^{\dagger} \mathsf{m}_1 \mathsf{\hat{x}} \\
		+ &\mathsf{J}(\mathsf{m}_1^{\dagger}\mathsf{m}_2+\mathsf{m}_1\mathsf{m}_2^{\dagger})+i \Psi \left(\mathsf{m}_1^{\dagger} - \mathsf{m}_1  \right).
	\end{aligned}
\end{equation}
Where $\Delta_{\mathsf{c}}=\omega_{\mathsf{c}}-\omega_0$, $\Delta_{\mathsf{m}_{1(2)}}=\omega_{\mathsf{m}_{1(2)}}-\omega_0$ are the detunings of the cavity mode and the two magnon modes respectively, with respect to the driving field. With each mode's dissipation and input noise taken into account, a set of Quantum Langevin Equations (QLEs) can be used to characterize the system's dynamical evolution as follows
\begin{align}
\frac{d\mathsf{c}}{dt}&=-(i\Delta_{\mathsf{c}}+\mathsf{k}_{\mathsf{c}})\mathsf{c}-i\mathsf{g}_1\mathsf{m}_1-i\mathsf{g}_2\mathsf{m}_2+\sqrt{2\mathsf{k}_{\mathsf{c}}}\mathsf{c}^{\rm{in}},\\ \nonumber
 \frac{d\mathsf{{m}}_1}{dt}&=-(i(\Delta_{\mathsf{m}_1}+\Delta_{B})+\mathsf{k}_{\mathsf{m}_1})\mathsf{m}_1-i\mathsf{G}_0\mathsf{m}_1\mathsf{\hat{x}}-i\mathsf{g}_1\mathsf{c}-i\mathsf{J}\mathsf{m}_2+\Psi+\sqrt{2\mathsf{k}_{\mathsf{m}_1}}\mathsf{m}_{1}^{\rm{in}},\\ \nonumber
   \frac{d\mathsf{m}_2}{dt}&=-(i\Delta_{\mathsf{m}_2}+\mathsf{k}_{\mathsf{m}_2})\mathsf{m}_2-i\mathsf{g}_2\mathsf{c}-i\mathsf{J}\mathsf{m}_1+\sqrt{2\mathsf{k}_{\mathsf{m}_2}}\mathsf{m}_{2}^{\rm{in}},\\ \nonumber
\frac{d\mathsf{\hat{x}}}{dt}&=\omega_{\mathsf{b}}\mathsf{\hat{y}},\\ \nonumber
\frac{d\mathsf{\hat{y}}}{dt}&=-\omega_{\mathsf{b}}\mathsf{\hat{x}}-\gamma_{\mathsf{b}}\mathsf{\hat{y}}-\mathsf{G}_0\mathsf{m}^{\dagger}_1\mathsf{m}_1+\Lambda,
\end{align} 
the effective cavity decay rate is represented by $\mathsf{k}_{\mathsf{c}}$. The dissipation rates for the two magnon modes and the phonon mode are given by $\mathsf{k}_{\mathsf{m}_{1(2)}}$ and $\gamma_{\mathsf{b}}$, respectively. The Rabi frequency $\Psi$ indicates that only magnon mode $\mathsf{m}_1$ is directly driven. The following correlation functions are separated into those with a zero mean
 \begin{equation}
 \begin{aligned} 
\langle \mathsf{c}^{\rm{in}}(t)\mathsf{c}^{\rm{in}\dagger}(t')\rangle =[\mathsf{N}_{\mathsf{c}}(\omega_{\mathsf{c}})+1]\delta(t-t'),&\quad
\langle \mathsf{c}^{\rm{in}}(t)\mathsf{c}^{\rm{in}\dagger}(t')\rangle =[\mathsf{N}_{\mathsf{c}}(\omega_{\mathsf{c}})]\delta(t-t').\\
\langle \mathsf{m}_{1(2)}^{\rm{in}}(t)\mathsf{m}_{1(2)}^{\rm{in}\dagger}(t')\rangle  =[\mathsf{N}_{\mathsf{m}_{1(2)}}(\omega_{\mathsf{m}_{1(2)}})+1]\delta(t-t'), &\quad \langle \mathsf{m}_{1(2)}^{\rm{in}\dagger}(t)\mathsf{m}_{1(2)}^{\rm{in}}(t')\rangle  =\mathsf{N}_{\mathsf{m}_{1(2)}}(\omega_{\mathsf{m}_{1(2)}})\delta(t-t').
\end{aligned}
\end{equation}
The Brownian motion of the mechanical mode is explained by the Langevin force operator $\Lambda$, which is autocorrelated as 
\begin{equation}
\frac{{\langle\Lambda(t)\Lambda(t')+\Lambda(t')\Lambda(t)\rangle}}{2\delta(t-t')}\approx\gamma_{\mathsf{b}}[2\mathsf{N}_{\mathsf{b}}(\omega_{\mathsf{b}})+1],
\end{equation}
 where we consider a high quality factor $\mathsf{Q}_{\mathsf{b}}=\omega_{\mathsf{b}}/\gamma_{\mathsf{b}}\gg 1$ for the mechanical oscillators to validate the Markov approximation \cite{vitali2007optomechanical}, with $\mathsf{N}_{\alpha}(\omega_{\alpha})=\left[\exp(\frac{\hbar\omega_{\alpha}}{\mathsf{k}_{\mathsf{B}}T})-1\right]^{-1}$ with $(\alpha=\mathsf{c},\mathsf{m}_{1(2)},\mathsf{b})$ are the thermal photon, magnon, and phonon numbers at equilibrium, respectively.  Here $T$ is the environmental temperature and  $\mathsf{k}_{\mathsf{B}}$ being the Boltzmann constant.
 
\section{Linearization and covariant matrix} \label{sec3}

The magnon mode $\mathsf{m}_1$ is driven by a strong external microwave field, and the cavity and the two magnon modes have high-amplitude beamsplitter interactions, represented by $|\mathsf{m}_{1\mathsf{s}}|, |\mathsf{m}_{2\mathsf{s}}|$ and $|\mathsf{c_s}|$. To linearize the system's dynamics around its steady-state values, we decompose each operator as the sum of its steady-state average and a quantum fluctuation operator, $\Phi= \Phi_{\mathsf{s}}+\delta\Phi$, where $(\Phi=\mathsf{c},\mathsf{m}_{1(2)},\hat{\mathsf{x}},\hat{\mathsf{y}})$. By disregarding minor variations in the second term, we can focus on the quantum fluctuations of the system's dynamics, which are of particular interest as we study the quantum correlation characteristics of the two magnon modes. The system quadratures' fluctuations are described by the linearized quantum Langevin equations (QLEs) $ [\delta \mathsf{X},\delta \mathsf{Y},\delta \mathtt{X}_{\mathsf{m}_1},\delta \mathtt{Y}_{\mathsf{m}_1},\delta \mathtt{X}_{\mathsf{m}_2},\delta \mathtt{Y}_{\mathsf{m}_2},\delta \mathsf{\hat{x}},\delta \mathsf{\hat{y}}]$, with $\delta \mathsf{X}=\frac{\delta \mathsf{c}+\delta \mathsf{c}^{\dagger}}{\sqrt{2}},\delta \mathsf{Y}=\frac{i\left(\delta \mathsf{c}^{\dagger}-\delta \mathsf{c}\right)}{\sqrt{2}},\delta \mathtt{X}_{\mathsf{m}_{1(2)}}=\frac{\delta \mathsf{m}_{1(2)}+\delta \mathsf{m}^{\dagger}_{1(2)}}{\sqrt{2}},\delta \mathtt{Y}_{\mathsf{m}_{1(2)}}=\frac{i\left(\delta \mathsf{m}^{\dagger}_{1(2)}-\delta \mathsf{m}_{1(2)}\right)}{\sqrt{2}}$, can be written in the following compact matrix.
\begin{equation}\label{eq5}
\frac{d\lambda(t)}{dt}=\mathsf{A}\lambda(t)+\chi(t),
\end{equation}
where $\lambda(t)=[\delta \mathsf{X}(t),\delta \mathsf{Y}(t),\delta \mathtt{X}_{\mathsf{m_1}}(t),\delta \mathtt{Y}_{\mathsf{m_1}}(t),\delta \mathtt{X}_{\mathsf{m_2}}(t),\delta \mathtt{Y}_{\mathsf{m_2}}(t),\delta \hat{\mathsf{x}}(t),\delta \hat{\mathsf{y}}(t)]^{T}$ is the vector of the quadrature fluctuations, and $\chi(t)=[\sqrt{2\mathsf{k}_{\mathsf{c}}}\mathsf{X}^{\rm{in}}(t),\sqrt{2\mathsf{k}_{\mathsf{c}}}\mathsf{Y}^{\rm{in}}(t),\sqrt{2\mathsf{k}_{\mathsf{m}_1}}\mathtt{X}_1^{\rm{in}}(t),\sqrt{2\mathsf{k}_{\mathsf{m}_2}}\mathtt{Y}_1^{\rm{in}}(t),\sqrt{2\mathsf{k}_{\mathsf{m}_2}}\mathtt{X}_2^{\rm{in}}(t),\sqrt{2\mathsf{k}_{\mathsf{m}_2}}\mathtt{Y}_2^{\rm{in}}(t),0,\Lambda(t)]^T$ is the vector of the input noises entering the system. The drift matrix $\mathsf{A}$ is provided by
\[
\mathsf{A}=
\begin{pmatrix}
-\mathsf{k}_{\mathsf{c}}&\Delta_{\mathsf{c}}&0&\mathsf{g}_1&0&\mathsf{g}_2&0&0\\
-\Delta_{\mathsf{c}}&-\mathsf{k}_{\mathsf{c}}&-\mathtt{g}_1&0&-\mathsf{g}_2&0&0&0\\
0&\mathsf{g}_1&-\mathsf{k}_{\mathsf{m}_1}&\tilde{\Delta}_{\mathsf{m}_1}+\Delta_{B}&0&\mathsf{J}&-\mathsf{G}&0\\
-\mathsf{g}_1&0&-\tilde{\Delta}_{\mathsf{m}_1}-\Delta_{B}&-\mathsf{k}_{\mathsf{m}_1}&-\mathsf{J}&0&0&0\\
0&\mathsf{g}_2&0&\mathsf{J}&-\mathsf{k}_{\mathsf{m}_2}&\Delta_{\mathsf{m}_2}&0&0\\
-\mathsf{g}_2&0&-\mathsf{J}&0&-\Delta_{\mathsf{m}_2}&-\mathsf{k}_{\mathsf{m}_2}&0&0\\
0&0&0&0&0&0&0&\omega_{\mathsf{b}}\\
0&0&0&\mathsf{G}&0&0&-\omega_{\mathsf{b}}&-\gamma_{\mathsf{b}}
\end{pmatrix},
\]
where $\tilde{\Delta}_{\mathsf{m_1}}={\Delta}_{\mathsf{m_1}}+ \mathsf{G}_0  \mathsf{\hat{x}_s}$ and the effective magnomechanical coupling rate, is represented by  $\mathsf{G}=i\sqrt{2}\mathsf{G}_{0} \mathsf{m}_{1\mathsf{s}}$. 
The average $\mathsf{c_s}$, $\mathsf{\hat{y}_s}$,$\mathsf{m}_{1(2)\mathsf{s}}$ and $\mathsf{\hat{x}_s}$  are given by
\begin{align}
\mathsf{c_s}&\simeq \frac{i\Psi(\mathsf{g}_1\Delta_{\mathsf{m}_2}-\mathsf{g}_2\mathsf{J})}{\Delta_{\mathsf{c}}(\tilde{\Delta}_{\mathsf{m}_1}+\Delta_B)\Delta_{\mathsf{m}_2}-\mathsf{J}^2\Delta_{\mathsf{c}}-\mathsf{g}^2_1\Delta_{\mathsf{m}_2}-\mathsf{g}_2^2(\tilde{\Delta}_{\mathsf{m}_1}+\Delta_B)+2\mathsf{J}\mathsf{g}_1\mathsf{g}_2},\\ \nonumber
\mathsf{m}_{1\mathsf{s}}&\simeq-\frac{\left(\mathsf{g}_1\Delta_{\mathsf{m}_2}-\mathsf{g}_2\mathsf{J}\right)\mathsf{c_s}+i\Delta_{\mathsf{m}_2}\Psi}{\Delta_{\mathsf{m}_2}(\tilde{\Delta}_{\mathsf{m}_1}+\Delta_B)-\mathsf{J}^2},\\ \nonumber
 \mathsf{m}_{2\mathsf{s}}&\simeq-\frac{\mathsf{g_2}\mathsf{c_s}+\mathsf{J}\mathsf{m_{1s}}}{\Delta_{\mathsf{m_2}}},\\ \nonumber
 \mathsf{\hat{y}_{s}}&\simeq 0,\\ \nonumber
 \mathsf{\hat{x}_{s}}&\simeq -\frac{\mathsf{G}_0}{\omega_{\mathsf{b}}}|\mathsf{m_{1s}}|^2,
\end{align}  
the expressions for $\mathsf{c_s}$ and $\mathsf{m}_{1(2)\mathsf{s}}$ are obtained under the condition that $|\Delta_{\mathsf{c}}|,|\tilde{\Delta}_{\mathsf{m}_1}|,|\Delta_{\mathsf{m}_2}|\gg \mathsf{k}_{\mathsf{c}},\mathsf{k}_{\mathsf{m}_{1(2)}}$ \cite{intro8}. In this instance, $\mathsf{c_s}$ and $\mathsf{m}_{1(2)\mathsf{s}}$ are pure imaginary numbers. Based on the Routh-Hurwitz criteria \cite{dejesus1987routh}, the system is stable. In this case, the drift matrix $\mathsf{A}$ is provided. The state of the system can be described in the stationary regime using an $8 \times 8$ covariance matrix (CM) $\mathsf{C}$ \cite{vitali2007optomechanical}, whose elements are written as:
\begin{equation} \label{CM}
\mathsf{C}_{\alpha\beta}= \frac{\langle\lambda_{\alpha} (\infty)\lambda_{\beta} (\infty) + \lambda_{\beta} (\infty)\lambda_{\alpha} (\infty)\rangle}{2}.
\end{equation}
Integrating Eq.~(\ref{eq5}) 
\begin{equation}
\label{QLEs2} 
\lambda(t)=\Psi(t)\lambda(0)+\int_{0}^{t}ds \Psi(r)\chi(t-r),
\end{equation}
with $\Psi(r)=\exp(\mathsf{A}r)$. If the stability of matrix $\mathsf{A}$ is satisfied, $\Psi(\infty)=0$, the Eq. (\ref{QLEs2}), writes as
\begin{equation}
\label{eq:uinf}
\lambda(\infty)=\displaystyle\lim_{t\to\infty} \int_{0}^{t}dr \,\Psi(r)\chi(t-r).
\end{equation}
The covariance matrix $\mathsf{C}_{\alpha\beta}$ which given by Eq. (\ref{CM}), is written as
\begin{equation}
\label{eq:corrmat}
\mathsf{C}_{\alpha\beta}=\sum_{\gamma,\gamma'}\int_{0}^{\infty} \int_{0}^{\infty}drdr'\Psi_{\alpha\gamma}(r)\Psi_{\beta\gamma'}(r')\Phi_{\gamma\gamma'}(r-r'),
\end{equation}
where $\Phi_{\gamma\gamma'}(r-r')=\frac{\langle \chi_{\gamma}(r)\chi_{\gamma'}(r')+ \chi_{\gamma'}(r)\chi_{\gamma}(r')\rangle}{2}$. For a large mechanical quality factor ($Q_{\phi}\gg 1$), thus 
$$\mathsf{F}_{\gamma\gamma'}=\frac{\Phi_{\gamma\gamma'}(r-r')}{\delta(r-r')},$$
where $\mathsf{F}_{\gamma\gamma'}$ is the diffusion matrix, describes the stationary noise correlations. The matrix $\mathsf{F}_{\gamma\gamma'}$ is defined by $\mathsf{F}_{\alpha\beta}=\frac{\langle \chi_{\alpha}(t)\chi_{\beta}(t')+\chi_{\alpha}(t')\chi_{\beta}(t)\rangle}{2\delta(t-t')}$, determined as $\mathsf{F}=\text{diag}[\mathsf{k}_{\mathsf{c}}(2\mathsf{N}_{\mathsf{c}}+1),\mathsf{k}_{\mathsf{c}}(2\mathsf{N}_{\mathsf{c}}+1),\mathsf{k}_{\mathsf{m}_1}(2\mathsf{N}_{\mathsf{m}_1}+1),\mathsf{k}_{\mathsf{m}_1}(2\mathsf{N}_{\mathsf{m}_1}+1),\mathsf{k}_{\mathsf{m}_2}(2\mathsf{N}_{\mathsf{m}_2}+1),\mathsf{k}_{\mathsf{m}_2}(2\mathsf{N}_{\mathsf{m}_2}+1),0,\gamma_{\mathsf{b}}(2\mathsf{N}_{\mathsf{b}}+1)]$. Then the Eq.~(\ref{eq:corrmat}) can be writes as
\begin{equation}
\label{eq:Cint}
\mathsf{C}=\int_{0}^{\infty}dr\Psi(r)\mathsf{F}\Psi(r)^{T}.
\end{equation}
The Lyapunov equation for the steady-state covariance matrix, represented by $\mathsf{C}$, can be obtained by integration by parts as follows
\begin{equation}
	\label{1n}
	\mathsf{A} \mathsf{C}+\mathsf{C} \mathsf{A}^T+\mathsf{F}=0.
\end{equation}
After obtaining the covariance matrix for the steady-state system from the previous calculations, we can investigate the bipartite and tripartite entanglement among the four modes. The state of a particular subsystem can be retrieved by eliminating the rows and columns corresponding to the uninteresting modes from the full covariance matrix $\mathsf{C}$. We then quantify Gaussian bipartite entanglement using logarithmic negativity and tripartite entanglement using the minimum residual contangle.

\section{Bipartite and Tripartite Entanglement}\label{sec4}
\subsection{Logarithmic negativity}

To quantify the bipartite entanglement, we use the logarithmic negativity $E_n$ \cite{Vidal, Plenio2005,Adesso1}, which is as follows
\begin{equation}\label{LN}
E_n=\max\left[0,-\ln(2\upsilon^-)\right],
\end{equation}  
the minimum symplectic eigenvalue of the $4\times4$ covariance matrix (CM) is characterized by $\upsilon^{-}=\min \text{eig}\left|\bigoplus_{j=1}^2(-\sigma_y)\tilde{C}_4\right|$. Here, $\tilde{C}_4$ is defined as $\rho_{1|2}C_{in}\rho_{1|2}$, where $C_{in}$ is a $4\times4$ matrix of any bipartition that can be generated by removing the uninteresting columns and rows in the full CM. The matrix $\rho_{1|2}=\text{diag}(1,-1,1,1)$ is used to perform a partial transposition at the CM level, and $\sigma_y$ is the $y$-Pauli matrix. Furthermore, the existence of entanglement between any bipartite subsystem is established by the condition $E_n>0$.

\subsection{Tripartite entanglement}

A minimal residual contangle, a measure of tripartite entanglement, is defined as \cite{Adesso1,Adesso2}
\begin{equation}
\rm{R}_{\tau}^{\text{min}}\equiv\min[\rm{R}_{\tau}^{a|bc},\rm{R}_{\tau}^{b|ac},\rm{R}_{\tau}^{c|ba}],
\end{equation}
the minimal residual contangle ($\rm{R}_{\tau}^{\text{min}}$) is a measure of quantum entanglement monogamy that ensures the tripartite entanglement remains invariant under all mode permutations. This makes it a true three-way property for any three-mode Gaussian state. It is given by
\begin{equation}
\rm{R}_{\tau}^{\alpha|\beta\gamma}=\mathcal{S}_{\alpha|\beta\gamma}-\mathcal{S}_{\alpha|\beta}-\mathcal{S}_{\alpha|\gamma}\geq 0,\quad \quad (\alpha,\beta,\gamma=a,b,c).
\end{equation}
The condition $\mathcal{S}_{\alpha|\beta\gamma}\geq\mathcal{S}_{\alpha|\beta}+\mathcal{S}_{\alpha|\gamma}$ is analogous to the Coffman-Kundu-Wootters monogamy inequality, which applies to a system of three qubits \cite{coffman} and denotes the existence of tripartite entanglement. The term $\mathcal{S}_{\alpha|\beta}$ is a proper entanglement monotone, which can be measured as $\mathcal{S}_{\alpha|\beta}=E_{\alpha|\beta}^2$. When calculating the logarithmic negativity between one and two modes, $E_{\alpha|\beta\gamma}$, one must adopt the definition of Eq. (\ref{LN}) by modifying the fundamental definition of $\upsilon^{-}$ as follows 
\begin{equation}
\upsilon^{-}=\min \text{eig}\left|\bigoplus_{j=1}^3(-\sigma_y)\tilde{C}_4\right|,
\end{equation}
where $\tilde{C}_4=\mathcal{K}C_4\mathcal{K}$, $\tilde{C}=\mathcal{K}_{i|jk}C\mathcal{K}_{i|jk}$, and the partial transpositionmatrices are given by
$\mathcal{K}_{1|23}=\text{\diag}(1,-1,1,1,1,1)~~,~~\mathcal{K}_{2|13}=\text{\diag}(1,1,1,-1,1,1)~~\text{and}~~\mathcal{K}_{3|12}=\text{\diag}(1,1,1,1,1,-1).$

\section{Results and Discusion} \label{sec6}

In this section, we will discuss our numerical results and delve into the evolution of quantum correlations within the system and its behavior when the Barnett effect is incorporated. We have employed experimentally attainable parameters \cite{zhang2016cavity,li2019entangling,hamza2024}, as in the table \ref{table}
\begin{table}[H]
\centering
\caption{Parameters used in numerical simulation.}
\begin{tabular}{l c c c}
\hline\hline
\textbf{Parameter} & \textbf{Symbol} &\textbf{Value} & \textbf{Unit} \\
\hline
Microwave mode frequency &$\omega_{\mathsf{c}}$                                     & $10^{10}\times 2\pi$          & Hz \\
Phonon mode frequency &$\omega_{\mathsf{b}}$                                        & $10^{10}\times 2\pi$          & Hz \\
Dissipation rate of phonon mode &$\gamma_{\mathsf{b}}$                              & $10^{2}\times 2\pi$            & Hz \\
Dissipation rate of magnon(1) mode &$\mathsf{k}_{\mathsf{m}_{1}}$                   & $10^{7}\times 2\pi$            & Hz \\
Dissipation rate of magnon(2) mode &$\mathsf{k}_{\mathsf{m}_{2}}$                   & $10^{7}\times 2\pi$            & Hz \\
Coupling rate magnon(1)-cavity mode &$\mathsf{g}_1$                                 & $3.2 \times 10^{6}\times 2\pi$ & Hz \\
Coupling rate magnon(2)-cavity mode &$\mathsf{g}_2$                                 & $2.6 \times 10^{6}\times 2\pi$ & Hz \\
Coupling rate magnon-magnon mode &$\mathsf{G}$                                      & $4.8 \times 10^{6}\times 2\pi$ & Hz \\
Temperature &$T$                                                                    & $10^{-2}$                     & K  \\
\hline\hline
\end{tabular}
\label{table}
\end{table}

\subsection{Bipartite and tripartite quantum correlations}
\begin{figure}[!h]
\includegraphics[width=0.33\linewidth]{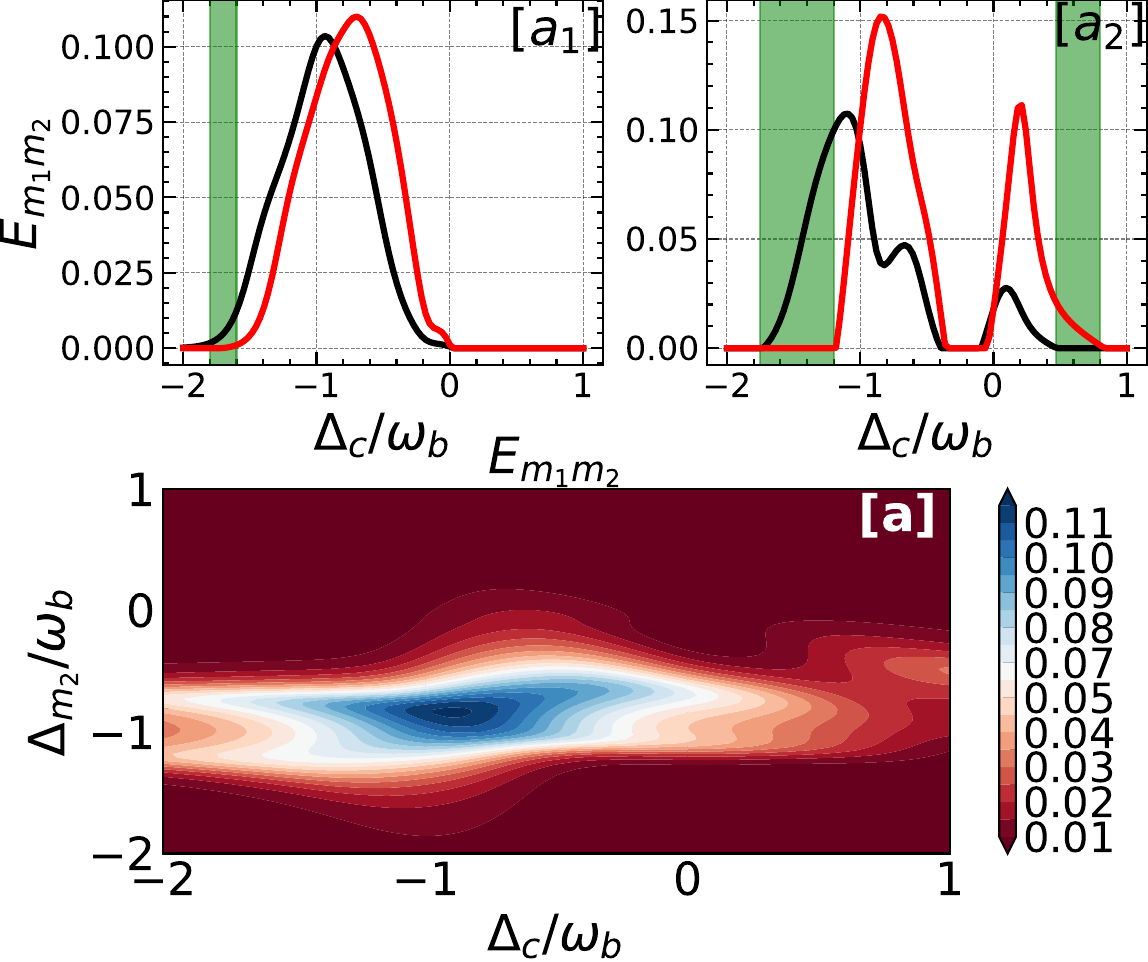} 
\includegraphics[width=0.33\linewidth]{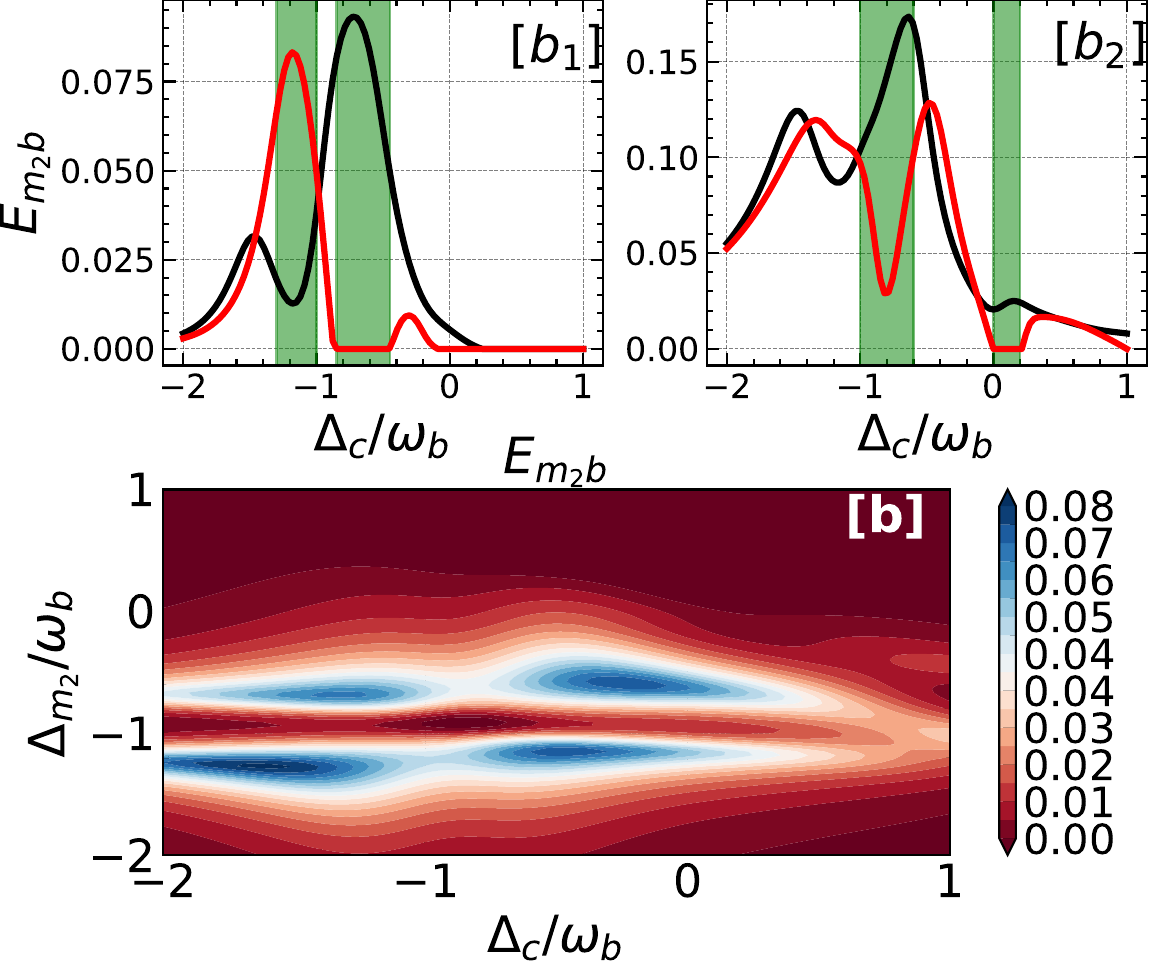}
\includegraphics[width=0.33\linewidth]{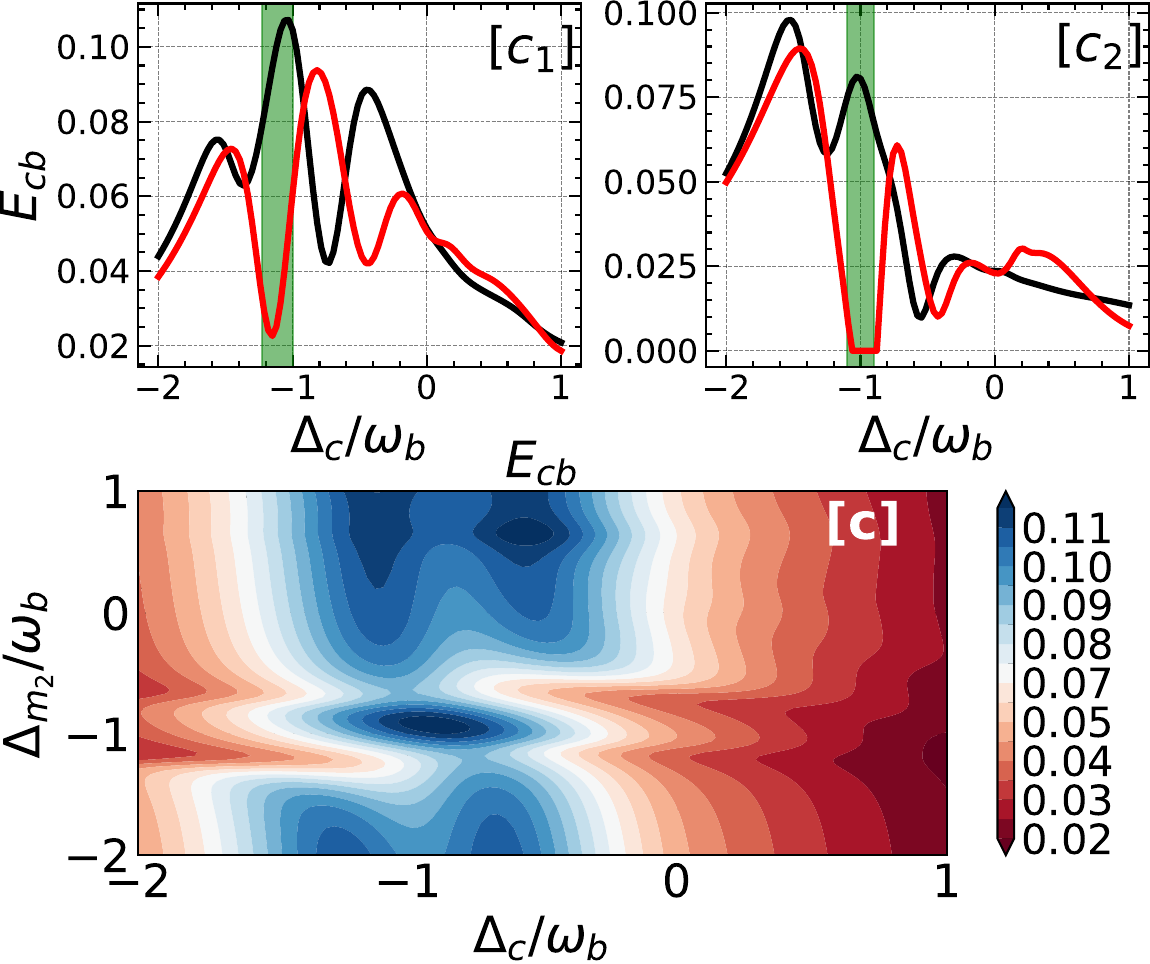}
\begin{center}
\caption{Plot of the logarithmic negativities between different bipartite: $([\rm a_1],[\rm a_2],[\rm a])$ $E_{\mathsf{m}_1\mathsf{m}_2}$, $([\rm b_1],[\rm b_2],[\rm b])$ $E_{\mathsf{m}_2\mathsf{b}}$, and $([\rm c_1],[\rm c_2],[\rm c])$ $E_{\mathsf{c}\mathsf{b}}$, versus the normalized detuning $\Delta_{\mathsf{c}}/\omega_{\mathsf{b}}$, the normalized magnetic field detuning $\Delta_B/\omega_{\mathsf{b}}$ and magnon-magnon coupling strength $\mathsf{J}$. In $([\rm a],[\rm b],[\rm c])$, we use $\mathsf{J}$ and $\Delta_B$ are 0. $\Delta_B > 0$ ($\Delta_B < 0$) corresponds to the magnetic field driven from the direction of $+z$ ($-z$). The green vertical strips in $([\rm a_{1,2}]-[\rm c_{1,2}])$ indicate regions of strong nonreciprocity. We select in $|\Delta_B|/\omega_{\mathsf{b}}=0.2$ (black line: $\Delta_B>0$ and red line: $\Delta_B<0$), with $\mathsf{J}=0$ in $([\rm a_{1}]-[\rm c_{1}])$ and $\mathsf{J}=\mathsf{g}_1$ in $([\rm a_{2}]-[\rm c_{2}])$. The other parameters are provided in the table \ref{table}.}
\label{fig:3}
\end{center}
\end{figure}

In Fig. \ref{fig:3} ([a], [b], and [c]), we explore the density plot of the bipartite entanglement for the three bipartitions: $E_{\mathsf{m}_1\mathsf{m}_2}$, $E_{\mathsf{m}_2\mathsf{b}}$, and $E_{\mathsf{c}\mathsf{b}}$. The plots are a function of the cavity detuning $\Delta_{\mathsf{c}}/\omega_{\mathsf{b}}$ and the magnon mode $\mathsf{m}_2$ detuning $\Delta_{\mathsf{m}_2}/\omega_{\mathsf{b}}$, with $\mathsf{J}=\Delta_{B}=0$. The bipartite entanglements $E_{\mathsf{m}_1\mathsf{m}_2}$ and $E_{\mathsf{c}\mathsf{b}}$ reach their maximum values around $\Delta_{\mathsf{c}}=-\omega_{\mathsf{b}}$ and when $\Delta_{\mathsf{m}_2}=-\omega_{\mathsf{b}}$. For a fixed value of $\Delta_{\mathsf{m}_1}$, as the cavity detuning increases, the entanglement for both magnon-magnon mode $E_{\mathsf{m}_1\mathsf{m}_2}$ and phonon-photon mode $E_{\mathsf{c}\mathsf{b}}$ increases until it reaches a maximum, after which it begins to decrease. In this case, the entanglement primarily arises from magnon-phonon interaction and is then gradually transferred to the cavity-magnon and magnon-magnon subsystems.

In Fig. \ref{fig:3} ([a$_1$], [b$_1$], and [c$_1$]), we plot all bipartite entanglements as a function of the cavity detuning $\Delta_{\mathsf{c}}$. We consider the case where the YIG spheres are spinning (i.e., in the presence of the Barnett effect), with $\mathsf{J}=0$ and $\Delta_B/\omega_{\mathsf{b}}=0.2$. In this scenario, the entanglement depends on the driving direction. As we have shown, there is a region where entanglement is created ($E_n\neq 0$) by driving from one side, while no entanglement ($E_n=0$) occurs when driving from the other. This region is ideal for generating nonreciprocal entanglement. 

In Fig. \ref{fig:3} ([a$_{2}$], [b$_{2}$], and [c$_{2}$]), we show the entanglement of the three bipartitions as a function of the cavity detuning $\Delta_{\mathsf{c}}$ for $\mathsf{J}=\mathsf{g}_1$ and $\Delta_B/\omega_{\mathsf{b}}=\pm 0.2$. The bipartitions $E_{\mathsf{m}_1\mathsf{m}_2}$ and $E_{\mathsf{m}_2\mathsf{b}}$ display enhanced entanglement due to the combined effects of the magnon-magnon coupling $\mathsf{J}$ and the Barnett effect, in contrast to $E_{\mathsf{c}\mathsf{b}}$, which does not exhibit such enhancement. This difference can be attributed to the strong magnon-phonon couplings, which effectively diminish the cavity-phonon interaction, thereby reducing $E_{\mathsf{c}\mathsf{b}}$ in comparison to the other bipartitions.

The maximum value of the magnon-magnon entanglement is significantly enhanced when considering the magnon-magnon coupling ($\mathsf{J}=\mathsf{g}_1$). Indeed, for $\mathsf{J}=0$ (Fig. \ref{fig:3} [a$_{1}$]), $E_{\mathsf{m}_1\mathsf{m}_2}\approx 0.11$ when $\Delta_{\mathsf{c}}/\omega_{\mathsf{b}}\approx -1.4$, whereas for $\mathsf{J}=\mathsf{g}_1$ (Fig. \ref{fig:3} [a$_{2}$]), $E_{\mathsf{m}_1\mathsf{m}_2}\approx 0.15$ around $\Delta_{\mathsf{c}}/\omega_{\mathsf{b}}\approx -1.4$. The green vertical strips indicate the regions of nonreciprocity. Nonreciprocal entanglement can be realized by tuning the angular frequencies $\Delta_B$, the detuning $\Delta_{\mathsf{c}}$, and the magnon-magnon coupling $\mathsf{J}$. This offers a promising strategy to enhance the performance of quantum devices by harnessing nonreciprocity.\\

\begin{figure}[!h]
\includegraphics[width=0.9\linewidth]{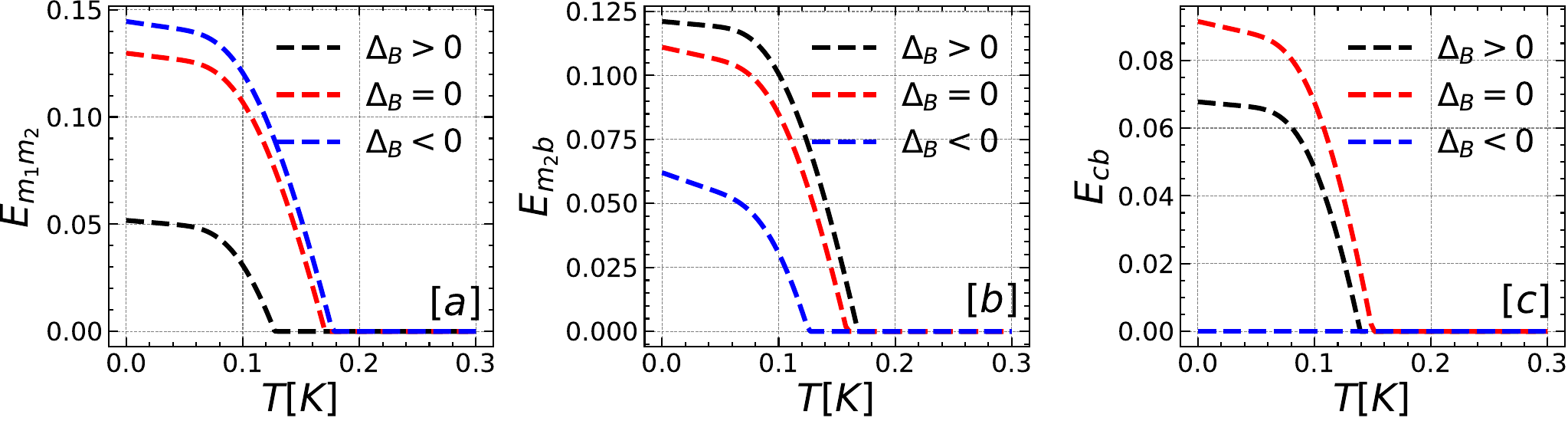} 
\begin{center}
\caption{Plot of the three bipartite entanglements magnon-magnon modes $E_{\mathsf{m}_1\mathsf{m}_2}$, magnon(2)-phonon modes $E_{\mathsf{m}_2\mathsf{b}}$ and photon-phonon modes $E_{\mathsf{c}\mathsf{b}}$ versus the temperature $T$ for various values of $\Delta_B=(0,\pm0.2\omega_{\mathsf{b}})$. The other parameters are provided in the table \ref{table}, with $\textcolor{red}{\mathsf{J}}/2\pi=3.2\times 10^6$ Hz.} 
\label{fig:4}
\end{center}
\end{figure}
We have previously demonstrated that, under the right conditions, bipartite entanglement can be made nonreciprocal. Here, as illustrated in Fig. \ref{fig:4}, we investigate how temperature affects bipartite entanglement for various values of $\Delta_{B}$ \cite{barnett}. We observe that the magnon-magnon mode entanglement, $E_{\mathsf{m}_1\mathsf{m}_2}$, is enhanced against temperature $T$ when $\Delta_B<0$, as shown in Fig. \ref{fig:4}[a]. The magnon(2)-phonon mode entanglement, $E_{\mathsf{m}_2\mathsf{b}}$, shows an enhancement with $\Delta_B>0$, as seen in Fig. \ref{fig:4}[b]. Additionally, the photon-phonon mode entanglement, $E_{\mathsf{c}\mathsf{b}}$, is strongest when the Barnett effect is null. When $\Delta_B<0$, the entanglement $E_{\mathsf{c}\mathsf{b}}$ is zero for all values of $T$. This implies that nonreciprocal entanglement can be made more robust against thermal noise by choosing appropriate parameters related to the Barnett effect. Furthermore, we understand that as the temperature decreases, bipartite entanglement becomes stronger. However, extremely low temperatures may also induce other undesired entanglements \cite{Vitali}. Therefore, one can slightly increase the temperature to completely eliminate entanglement in one direction at the expense of slightly decreasing it in the other, in order to achieve the ideal nonreciprocity of entanglement. This suggests that the system can always be made to exhibit optimal entanglement nonreciprocity by selecting a suitable temperature $T$. 

Figures \ref{fig:4}[a]-[c] clearly show that nonreciprocal magnon-magnon, magnon(2)-phonon, and photon-phonon entanglement can still be achieved for temperatures up to $T=0.18$K, $T=0.15$K, and $T=0.13$K, respectively. Previous works, however, reported a maximum temperature limit of $T=0.2$K for nonreciprocal entanglement realization \cite{chen}. Our findings suggest that when the Barnett effect is coupled with CMM devices, nonreciprocal entanglement exhibits exceptional robustness against bath temperature compared to previous investigations.

\begin{figure}[!h]
\includegraphics[width=0.9\linewidth]{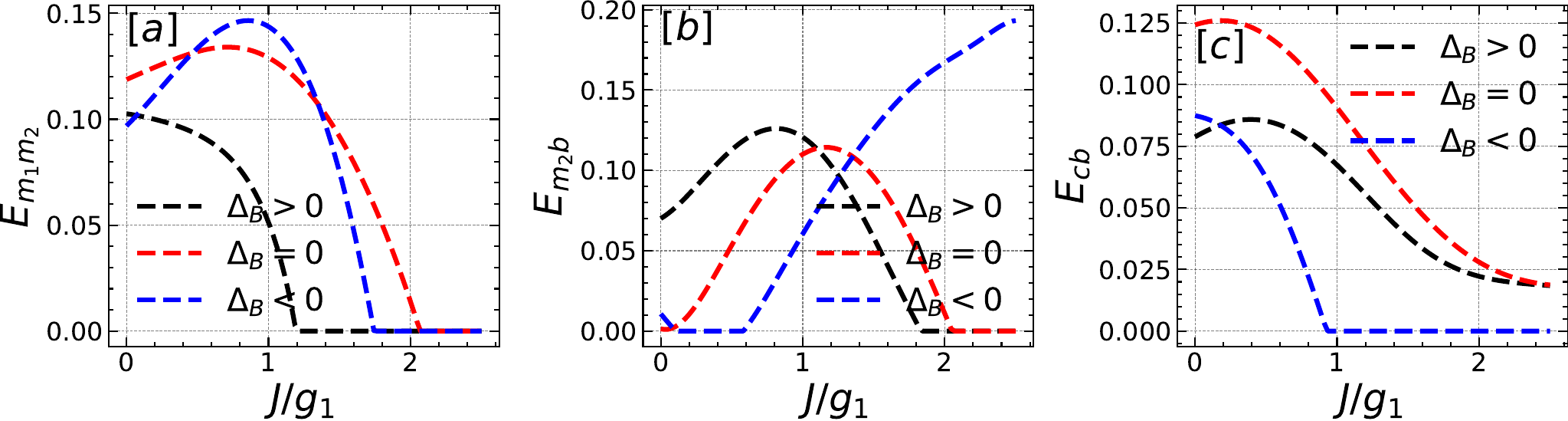} 
\begin{center}
\caption{Plot of the three bipartite entanglements magnon-magnon modes $E_{\mathsf{m}_1\mathsf{m}_2}$, magnon(2)-phonon modes $E_{\mathsf{m}_2\mathsf{b}}$ and photon-phonon modes $E_{\mathsf{c}\mathsf{b}}$ versus the magnon-magnon coupling strength $\mathsf{J}/\mathsf{g}_1$ for various values of $\Delta_B$. The other parameters are provided in the table \ref{table}.} 
\label{fig:5}
\end{center}
\end{figure}

In Fig. \ref{fig:5}, we plot the variation of the three bipartite entanglements magnon-magnon modes ($E_{\mathsf{m}_1\mathsf{m}_2}$), magnon(2)-phonon modes ($E_{\mathsf{m}_2\mathsf{b}}$), and photon-phonon modes ($E_{\mathsf{c}\mathsf{b}}$) as a function of the magnon-magnon coupling strength $\mathsf{J}/\mathsf{g}_1$ for different values of $\Delta_B$. We observe that $E_{\mathsf{m}_1\mathsf{m}_2}$ exists when the coupling strength $\mathsf{J}=0$ for all three cases of the Barnett parameter, $\Delta_B$. While the entanglement for the cases $\Delta_B=0$ and $\Delta_B<0$ increases at low values of $\mathsf{J}$, the entanglement for the case $\Delta_B>0$ decreases and disappears when $\mathsf{J}>\mathsf{g}_1$. On the other hand, the entanglement for $E_{\mathsf{m}_1\mathsf{m}_2}$ increases for $\Delta_B=0$ and $\Delta_B<0$, but it vanishes when $\mathsf{J}>2\mathsf{g}_1$ and $\mathsf{J}>1.9\mathsf{g}_1$, respectively, as shown in Fig. \ref{fig:5}[a]. In Fig. \ref{fig:5}[b], we show the entanglement $E_{\mathsf{m}_2\mathsf{b}}$. It is zero for $\mathsf{J}=0$ and $\mathsf{J}>2\mathsf{g}_1$ when $\Delta_B=0$. The entanglement also increases for the case where $\Delta_B<0$. As illustrated in Fig. \ref{fig:5}[c], the entanglement for the photon-phonon mode decreases when $\Delta_B<0$ and becomes zero when $\mathsf{J}>\mathsf{g}_1$. For the other cases ($\Delta_B=0$ and $\Delta_B>0$), the entanglement remains non-zero for all values of $\mathsf{J}$.

To investigate how the Barnett effect influences nonreciprocal two-mode squeezing, we present the Wigner function is steady-state as 
\begin{equation}
W(\lambda_j) = \frac{1}{\sqrt{(2\pi)^2 \det[\tilde{C}_4]}} \exp\left [ -\frac{1}{2} \lambda_j^T \tilde{C}_4^{-1} \lambda_j \right ],
\end{equation}
where $\lambda_j$ is the column vector of the photon, magnon(1), magnon(2) and phonon mode fluctuations and $\tilde{C}_4$ stands for the covariance matrix for the each mode. 

\begin{figure}[!h]
\includegraphics[width=0.9\linewidth]{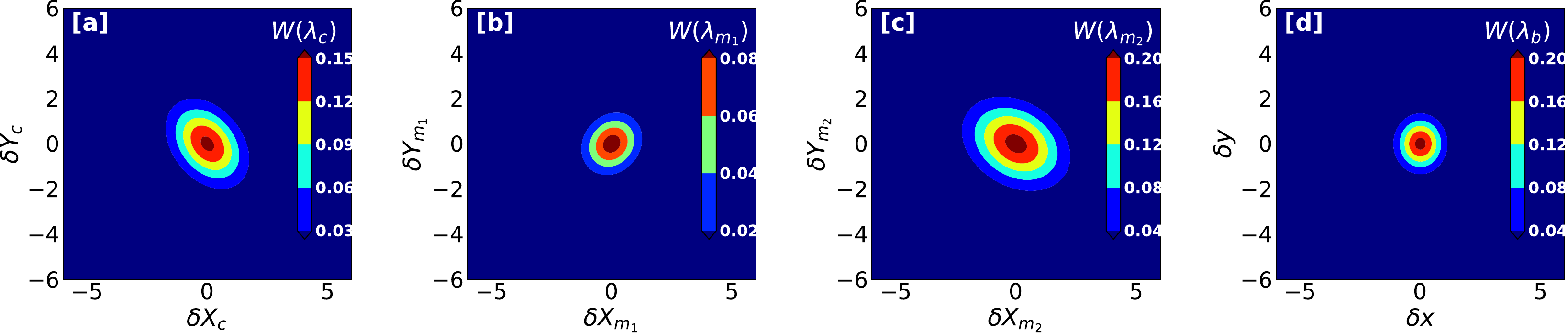}
\includegraphics[width=0.9\linewidth]{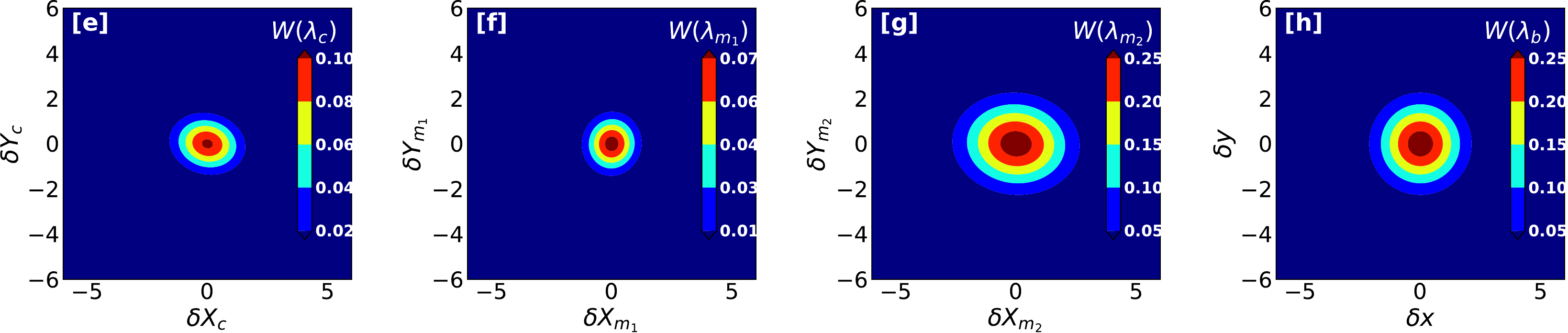} 
\begin{center}
\caption{The Wigner function $W(\rm{u})$ of the cavity mode $\mathsf{c}$ (a,e), the magnon(1) mode $\mathsf{m}_1$ (b,f), the magnon(2) mode $\mathsf{m}_2$ (c,g), and the phonon mode $\mathsf{b}$ (d,h) for $\Delta_B=0.2\omega_{\mathsf{b}}>0$ (a-d) and $\Delta_B=-0.2\omega_{\mathsf{b}}<0$ (e-h), respectively. The other parameters are provided in the table \ref{table}, with $\mathsf{J}/2\pi=3.2\times 10^6$ Hz.} 
\label{fig:6}
\end{center}
\end{figure}

In Fig. \ref{fig:6}, we plot the variation of the Wigner function $W(\rm{u})$ for the cavity mode $\mathsf{c}$ ([a, e]), the magnon(1) mode $\mathsf{m}_1$ ([b, f]), the magnon(2) mode $\mathsf{m}_2$ ([c, g]), and the phonon mode $\mathsf{b}$ ([d, h]). These plots are shown as a function of the cross-quadrature pairs ($\delta X_{\mathsf{c}},\delta Y_{\mathsf{c}}$) [a, e], ($\delta X_{\mathsf{m}_{1}},\delta Y_{\mathsf{m}_{1}}$) [b, f], ($\delta X_{\mathsf{m}_{2}},\delta Y_{\mathsf{m}_{2}}$)[c, g], and ($\delta \mathsf{x},\delta \mathsf{y}$)[d, h]. We consider different values of $\Delta_B$, where $\Delta_B>0$ ($\Delta_B<0$) corresponds to the magnetic field being driven from the $+z$ ($-z$) direction. The dashed circle and solid ellipse indicate a drop to $1/e$ of the maximum value of $W(\rm{u})$ for the vacuum and steady states, respectively. When $\Delta_B>0$, we observe an elliptical shape for the Wigner function $W(\rm{u})$. However, when $\Delta_B<0$, $W(u)$ decreases and its shape changes, as shown in Fig. \ref{fig:6}[a, e]. The Wigner function of the magnon(1) mode $\mathsf{m}_1$ is very low for both $\Delta_B>0$ and $\Delta_B<0$, and it retains a circular shape, as seen in Fig. \ref{fig:6}[b, f]. The Wigner function for the magnon(2) mode $\mathsf{m}_2$ is very large and equal to that of the phonon mode (Fig. \ref{fig:6}[d, h]), while the plots in Fig. \ref{fig:6}[c, g] show an elliptical shape for the Wigner function.\\

\begin{figure}[!h]
\includegraphics[width=0.33\linewidth]{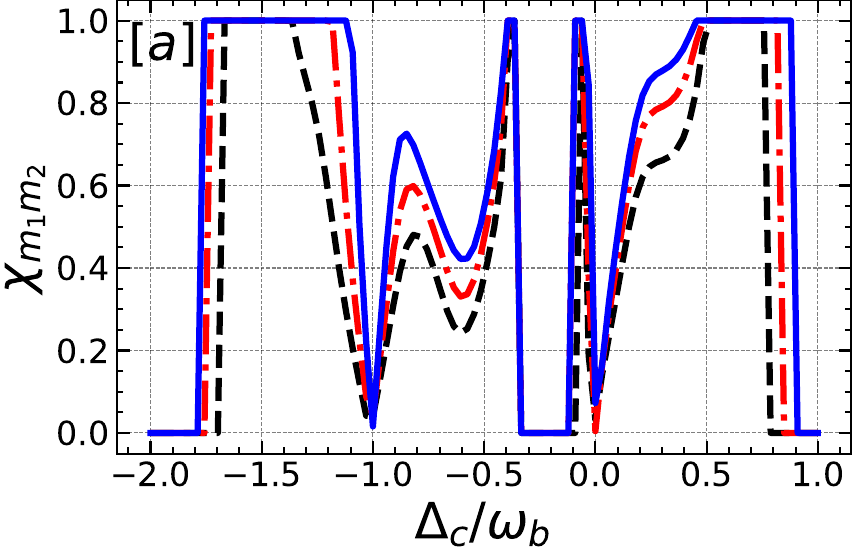}
\includegraphics[width=0.33\linewidth]{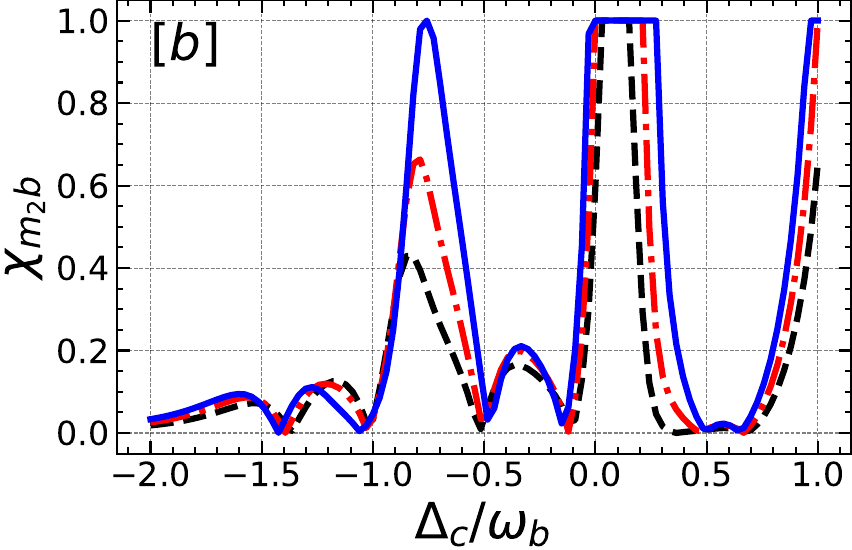} 
\includegraphics[width=0.33\linewidth]{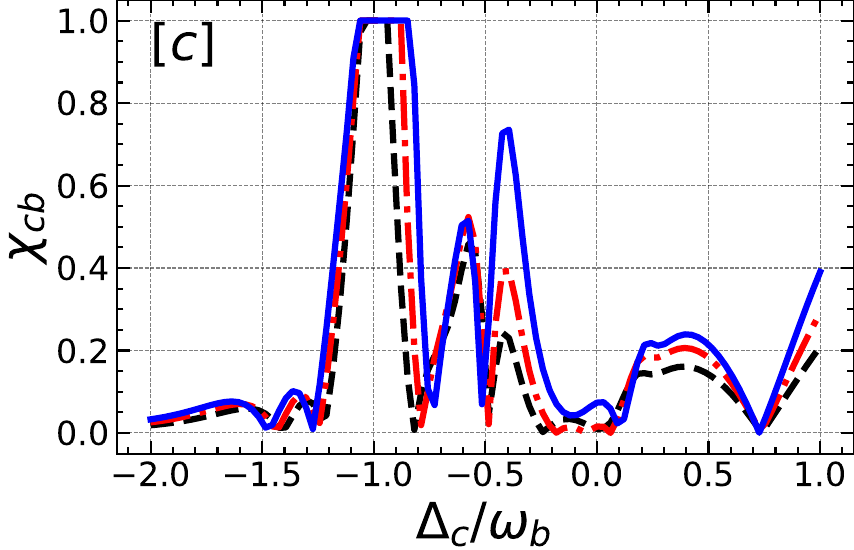} 
\begin{center}
\caption{The nonreciprocal of entanglement between magnon-magnon $E_{\mathsf{m}_1\mathsf{m}_2}$, magnon(2)-phonon modes $E_{\mathsf{m}_2\mathsf{b}}$ and photon-phonon modes $E_{\mathsf{c}\mathsf{b}}$ as a function of normalized detunings $\Delta_{\mathsf{c}}/\omega_{\mathsf{b}}$ for different Barnett effect $\Delta_B$ values, such as (black curve) $|\Delta_B|=0.15\omega_{\mathsf{b}}$, (red curve) $|\Delta_B|=0.20\omega_{\mathsf{b}}$ and (blue curve) $|\Delta_B =0.25\omega_{\mathsf{b}}$. Other parameters are the same as figure \ref{fig:6}, with $\mathsf{J}/2\pi=3.2\times 10^6$ Hz.}
\label{fig:7}
\end{center}
\end{figure}

We use the bidirectional contrast ratio $\mathcal{X}$ (which satisfies $0\leq \mathcal{X}\leq 1$) to quantitatively evaluate nonreciprocity of bipartite and tripartite entanglements \cite{barnett,chabar2025,hou2025,liu2025}
\begin{equation}
\label{contrast}
\mathcal{X}_{k l}=\frac{\left|E_{k l}\left(\Delta_B>0\right)-E_{k l}\left(\Delta_B<0\right)\right|}{E_{k l}\left(\Delta_B>0\right)+E_{k l}\left(\Delta_B<0\right)},
\end{equation}
\begin{equation}
\label{contrast2}
\mathcal{X}_{\rm{R}}=\frac{\left|\rm{R}_{\rm{min}}\left(\Delta_B>0\right)-\rm{R}_{\rm{min}}\left(\Delta_B<0\right)\right|}{\rm{R}_{\rm{min}}\left(\Delta_B>0\right)+\rm{R}_{\rm{min}}\left(\Delta_B<0\right)},
\end{equation}
a contrast ratio of $\mathcal{X}=0$ indicates no nonreciprocity, while a value of $\mathcal{X}=1$ corresponds to ideal nonreciprocity for bipartite entanglement.

Figure \ref{fig:7}[a] demonstrates that the nonreciprocity of bipartite entanglement can be switched on and off by tuning the auxiliary cavity detuning, $\Delta_{\mathsf{c}}$. The bidirectional contrast ratio for two types of entanglement can be tuned between 0 and 1 by varying $\Delta_{\mathsf{c}}$. Specifically, nonreciprocal entanglement occurs when $-1.9\leq\Delta_{\mathsf{c}}\leq-1.0$ and $0\leq\Delta_{\mathsf{c}}\leq0.8$. Figure \ref{fig:7}[b] shows how adjusting the auxiliary cavity frequency detuning can achieve optimal nonreciprocal magnon(2)-phonon and photon-phonon entanglements, as seen in Figure \ref{fig:7}[c]. In our scheme, optimal nonreciprocity is achieved when $-1.0\leq\Delta_{\mathsf{c}}/\omega_{\mathsf{b}}\leq-0.5$ and $-1.4\leq\Delta_{\mathsf{c}}/\omega_{\mathsf{b}}\leq-0.3$. This suggests that as the frequency detuning increases ($\Delta_{\mathsf{c}}>-1.5\omega_{\mathsf{b}}$), the entangled states become more distinct and asymmetric in their behavior in different directions. The nonreciprocity vanishes near $\Delta_{\mathsf{c}}/\omega_{\mathsf{b}}\approx0.68$, at which point the entangled states are indistinguishable in both directions. The ability to tune these two types of entanglement on and off through the bidirectional contrast ratio confirms a nonreciprocal process in the generation of entanglement in this system.

\begin{figure}[!h]
\begin{center}
\includegraphics[width=0.4\linewidth]{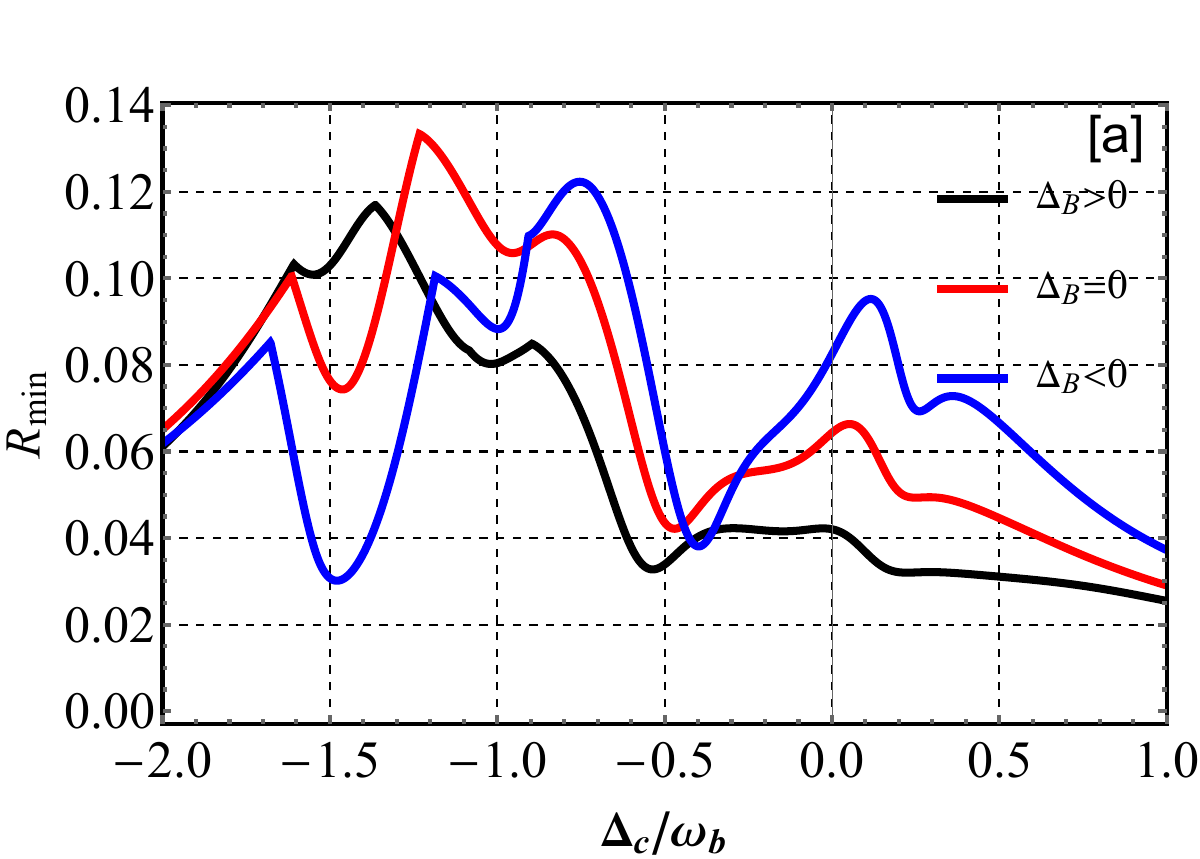} 
\includegraphics[width=0.4\linewidth]{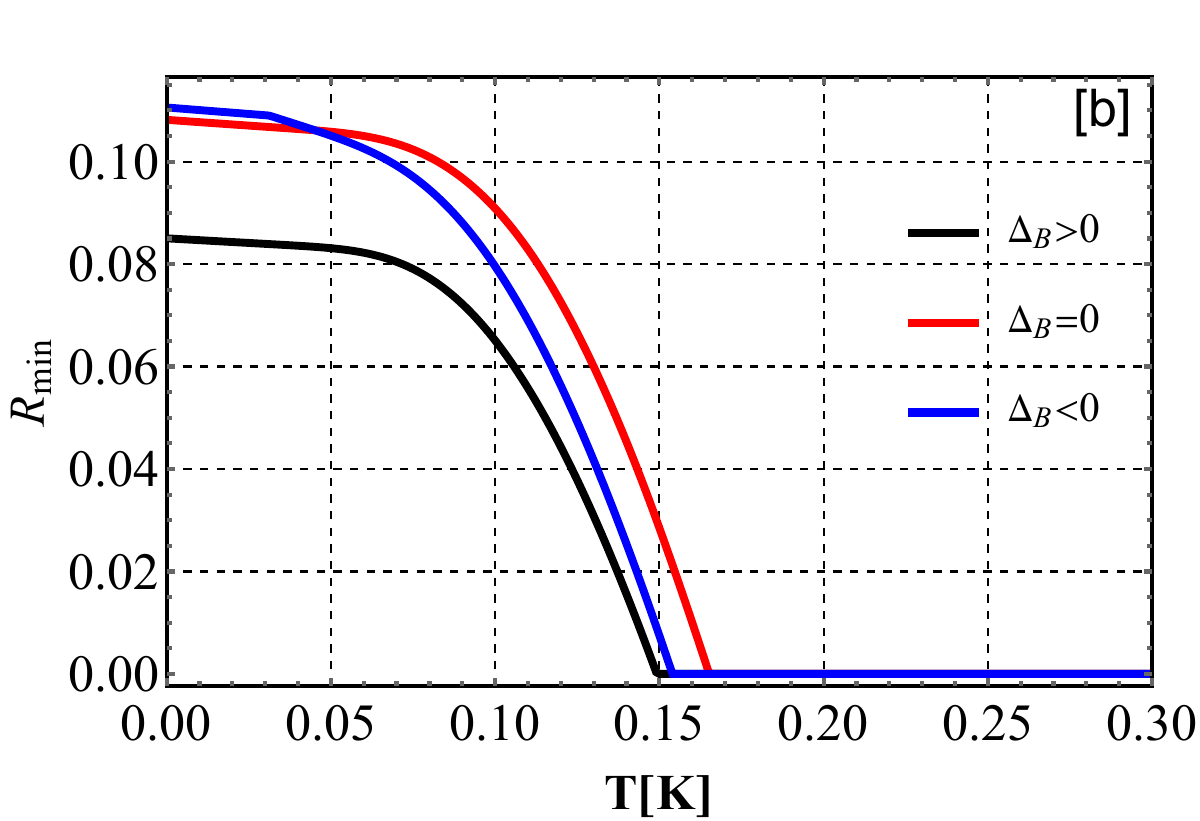}\\
\includegraphics[width=0.4\linewidth]{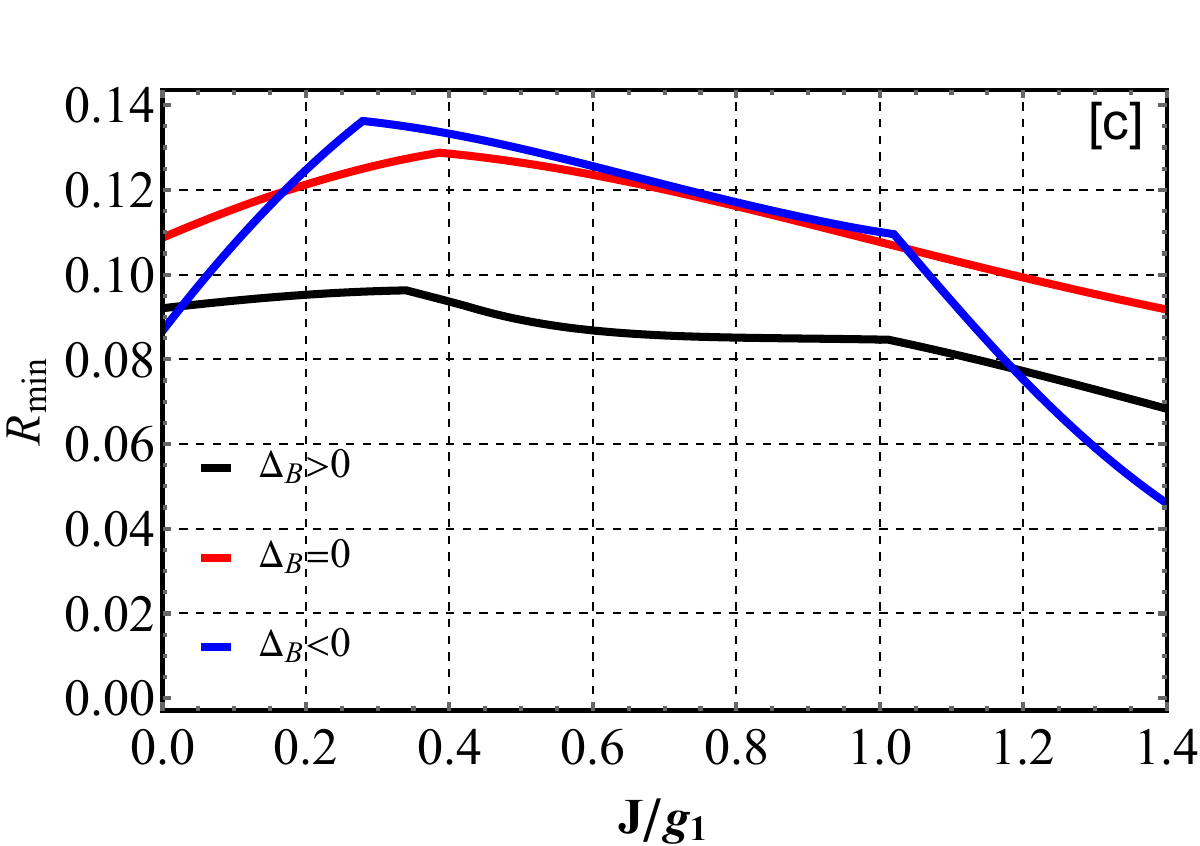}
\includegraphics[width=0.4\linewidth]{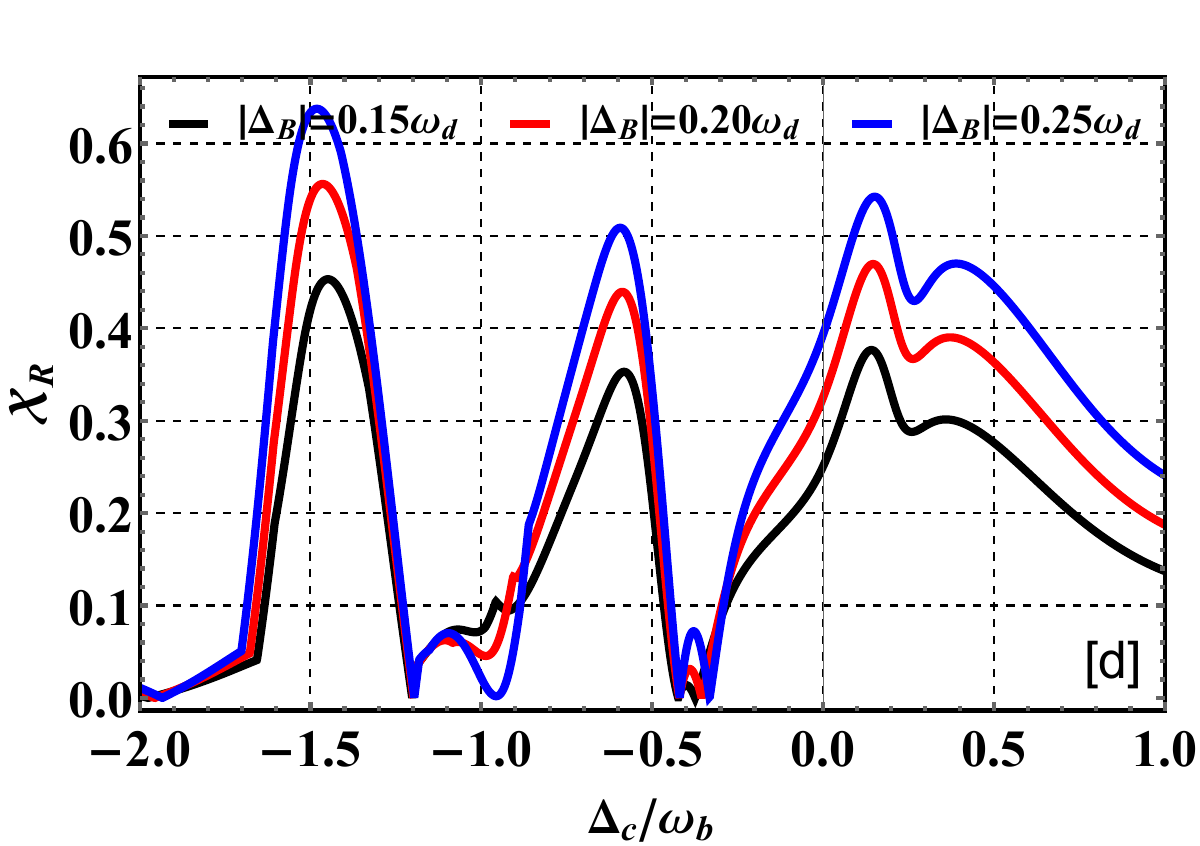}
\caption{Plot of the tripartite magnon(1)-photon-phonon ($\mathsf{m}_1-\mathsf{c}-\mathsf{b}$) entanglement $\rm{R}_{min}$ versus, such as [a] normalized detuning $\Delta_{\mathsf{c}}/\omega_{\mathsf{b}}$, [b] temperature $T$ and [c] normalized magnon-magnon coupling $\mathsf{J}/\mathsf{g}_1$ for different cases with respect to the Barnett effect effect $\Delta_B$, such as (black curve) $\Delta_B=0.2\omega_{\mathsf{b}}>0$, (red curve) $\Delta_B=0$ and (blue curve) $\Delta_B<0$. [d] The nonreciprocal of entanglement between magnon(1)-photon-phonon versus normalized detuning $\Delta_{\mathsf{c}}/\omega_{\mathsf{b}}$ for various value of the Barnett effect $\Delta_B$, such as (black curve) $|\Delta_B|=0.15\omega_{\mathsf{b}}$, (red curve) $|\Delta_B|=0.20\omega_{\mathsf{b}}$ and (blue curve) $|\Delta_B|=0.25\omega_{\mathsf{b}}$. The other parameters are provided in the table \ref{table}, with $\mathsf{J}/2\pi=3.2\times 10^6$ Hz.} 
\label{fig:8}
\end{center}
\end{figure}

In Fig. \ref{fig:8}, we investigate the variation of the tripartite entanglement between the magnon(1)-photon-phonon ($\mathsf{m}_1\mathsf{c}\mathsf{b}$) modes, measured by the minimal residual contangle, $\rm{R}_{min}$. We examine its behavior as a function of [a] the normalized detuning $\Delta_{\mathsf{c}}/\omega_{\mathsf{b}}$, [b] the temperature $T$, and [c] the normalized magnon-magnon coupling $\mathsf{J}/\mathsf{g}_1$ for different cases related to the Barnett effect, $\Delta_B$.\\
Fig. \ref{fig:8}[a] shows that the minimal residual contangle, $\rm{R}_{min}$, is enhanced when $\Delta_{\mathsf{c}}/\omega_{\mathsf{b}}<0$ for $\Delta_B=0$. For $\Delta_{\mathsf{c}}/\omega_{\mathsf{b}}>-0.5$, we also see an enhancement in the tripartite entanglement.
From Fig. \ref{fig:8}[b], we notice that the minimal residual contangle decreases as the temperature $T$ increases. When $\Delta_B=0$, the tripartite entanglement persists until $T>0.15$K. In contrast, for $\Delta_B>0$ and $\Delta_B<0$, the entanglement disappears when $T=0.15$K.
As shown in Fig. \ref{fig:8}[c], the tripartite entanglement between the magnon(1)-photon-phonon modes increases with an increase in the normalized magnon-magnon coupling $\mathsf{J}/\mathsf{g}_1$. However, when $\mathsf{J}>\mathsf{g}_1$, the minimal residual contangle, $\rm{R}_{min}$, begins to decrease.
Lastly, Fig. \ref{fig:8}[d] shows the variation of nonreciprocal entanglement between the magnon(1)-photon-phonon modes as a function of the normalized detuning $\Delta_{\mathsf{c}}/\omega_{\mathsf{b}}$ for various values of the Barnett effect, $\Delta_B$. We observe that the nonreciprocal entanglement varies periodically, and its magnitude is greater for $|\Delta_B|=0.25\omega_{\mathsf{b}}$ compared to $|\Delta_B|=0.20\omega_{\mathsf{b}}$ and $|\Delta_B|=0.15\omega_{\mathsf{b}}$. This demonstrates that increasing the Barnett effect contributes to the enhancement of nonreciprocal entanglement.

\section{Conclusion}  \label{sec7}

In summary, we investigated the enhancement of bipartite and tripartite entanglement in a cavity magnomechanical system through the Barnett effect. We discovered that by rotating the YIG sphere, the Barnett effect induces a Barnett frequency shift. We used logarithmic negativity to quantify the entanglement between different modes in the steady-state. The genuine tripartite entangled state was quantified using the minimum residual contangle. We showed that the Barnett effect can be used to generate both bipartite and tripartite nonreciprocal entanglement. While we found that both bipartite and tripartite quantum correlations are fragile under thermal noise. We have showed that the Barnett effect enhances entanglement under thermal effects and have generated squeezed states for the two magnon modes and the photon mode. Moreover, we have shown that magnon-magnon coupling enhances the entanglement between different two modes.

\end{document}